\documentclass[MSNbibl,nameyear,dvips]{arxstspdf}
\usepackage{graphicx}
\usepackage{flushend}
\usepackage{stfloats}

\volume{26}
\issue{1}
\pubyear{2011}
\firstpage{62}
\lastpage{83}
\doi{10.1214/10-STS343}

\makeatletter
\newproclaim{definition}[theorem]{Definition}

\newproclaim{example}{Example}
\newtheorem{theorem}{Theorem}
\newtheorem{proposition}{Proposition}
\newcommand{\remove}[1]{}
\newcommand{\p}[1]{{\operatorname{Pr}}(#1)}
\newcommand{\ms}[3]{\{ #1_i \}_{i=#2}^{#3}}
\newcommand{\beq}{\begin{eqnarray}}
\newcommand{\beqs}{\begin{eqnarray*}}
\newcommand{\eeq}{\end{eqnarray}}
\newcommand{\eeqs}{\end{eqnarray*}}
\newcommand{\ajtheta}{\theta\cdot a_j}
\newcommand{\E}{\mathrm{E}}

\makeatother

\begin{document}
\begin{frontmatter}

\title{Statistical Modeling of RNA-Seq Data}
\runtitle{Statistical Modeling of RNA-Seq Data}

\begin{aug}
\author{\fnms{Julia} \snm{Salzman}\thanksref{t1}\ead[label=e1]{julia.salzman@stanford.edu}},
\author{\fnms{Hui} \snm{Jiang}\corref{}\thanksref{t1}\ead[label=e2]{jiangh@stanford.edu}}
\and
\author{\fnms{Wing Hung} \snm{Wong}\ead[label=e3]{whwong@stanford.edu}}

\runauthor{J. Salzman, H. Jiang and W. H. Wong}

\affiliation{Stanford University}
\address{Julia Salzman is Research Associate, Department of
Statistics and
Biochemistry, Stanford University, Stanford, California 94305, USA
\printead{e1}. Hui Jiang is Postdoctoral
Scholar, Department of Statistics and Stanford Genome Technology
Center, Stanford University, Stanford, California 94305, USA
\printead{e2}. Wing Hung Wong is Professor of
Statistics and of Health Research and Policy, Stanford University,
Stanford, California 94305, USA \printead{e3}.}

\thankstext{t1}{These authors contributed equally to this work and are
both corresponding authors.}

\end{aug}

% ABSTRACT
%
\begin{abstract}
Recently, ultra high-throughput sequencing of RNA (RNA-Seq) has been
developed as an approach for analysis of gene expression. By obtaining
tens or even hundreds of millions of reads of transcribed sequences, an
RNA-Seq experiment can offer a comprehensive survey of the population of
genes (transcripts) in any sample of interest. This paper introduces a
statistical model for estimating isoform abundance from RNA-Seq data and
is flexible enough to accommodate both single end and paired end RNA-Seq
data and sampling bias along the length of the transcript. Based on the
derivation of minimal sufficient statistics for the model, a
computationally feasible implementation of the maximum likelihood
estimator of the model is provided. Further, it is shown that using paired
end RNA-Seq provides more accurate isoform abundance estimates than single
end sequencing at fixed sequencing depth. Simulation studies are also given.

%In mammalian cells, RNA molecules can have highly
%similar sequences yet can encode proteins with different functional
%roles. Isoforms of a gene are an
%example of a collection of such RNA sequences. Accumulating evidence
%suggests that a key factor
%characterizing cell function in mammals is
%differential isoform expression. Quantifying differences in cellular
%abundance of isoforms is therefore of significant biological interest.
% Ultra High Throughput Sequencing (UHTS) is an emerging technology
%which promises to become as (or more) powerful, popular and
%cost-effective than current microarray technology for estimating
%gene expression, particularly at the level of isoforms. Yet,
%statistical methods for performing such estimation are not well
%developed. This paper introduces a statistical model for
%estimating isoform abundance from UHTS data. The model is
%analyzed in detail, including derivation of minimal sufficient
%statistics and a study of the Fisher information. Further, a
%computationally feasible implementation of the maximum likelihood
%estimator of the model is provided.
\end{abstract}

% KEYWORDS
%

\end{frontmatter}

%s1 ###
\section{Introduction}

%s1.1 ###
\subsection{Biological Background}
All cells in an individual mammal have almost identical DNA. Yet, cell
function within an organism has huge
variation. One mechanism that differentiates cell function is its gene
expression pattern. Recent research
has shown that this differentiation may be on a fine scale: that subtle
sequence variants of expressed genes (also referred to as transcripts),
called isoforms, have significant impact on the function of the
proteins encoded by the RNA and hence their
function in the cell (see, e.g., \cite{Wang2008}). The purpose of this
paper is to develop and analyze
statistical methodology for measuring differential expression of
isoforms using an emerging powerful
technology called Ultra High Throughput Sequencing (UHTS). Such study
has the potential to help reveal new
insights into cellular isoform level gene expression patterns and
mechanisms, including characteristics of
cell specific specialization.

%f1 ###
\begin{figure}

\includegraphics{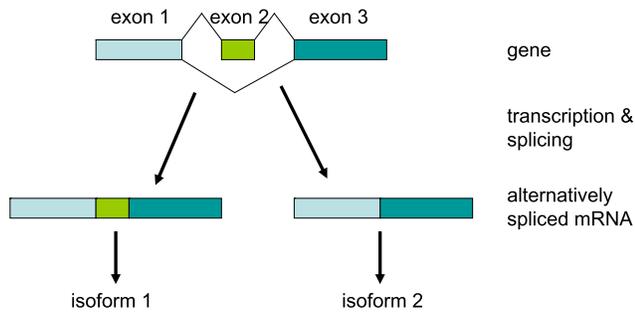}

\caption[Alternative Splicing]{A gene (DNA sequence) with three exons.
During transcription, two isoforms are generated. The first
isoform contains all of the gene's three exons. The second isoform
contains the first and third exon, skipping the middle exon. This
process is called alternative splicing and the middle exon is
called an alternatively spliced exon.}\label{fig:alternative_splicing}
\end{figure}

The central dogma in biology describes the information transfer that
allows cells to generate proteins, the building blocks of biological
function. This dogma states that DNA is transcribed to messenger RNA
(mRNA) which is in turn translated into proteins. Recent discoveries have
highlighted the importance of regulation at the level of mRNA, showing
that protein levels and function can be regulated by subtle differences in
the sequence of mRNA molecules in a cell.

In bacteria, short DNA sequences are transcribed in a one to one
fashion to
mRNA. This mRNA is referred to as a~gene or a transcript. Like DNA, each
mRNA is a
string of nucleotides, each position taking four possible values.
Mammalian cells commonly generate a large class of mRNA molecules from
a~single relatively short DNA sequence. The set of such mRNA molecules
are called isoforms of a~gene. This paper concentrates on one common
mechanism generating isoforms called alternative splicing. An example
of alternative splicing is depicted in Figure \ref{fig:alternative_splicing}: two isoforms can arise
from the same gene when
the DNA, which is comprised of three sequence blocks (called exons),
can be
transcribed into two different mRNA molecules: one of which contains
all three exons and
one of which only contains the first and third exon. As this example
shows, isoforms typical\-ly have highly similar
sequence. Despite this sequen\-ce similarity, isoforms can encode
proteins which may have different functional roles. Further, most~%
ge\-nes have more than three exons, and alternative use of exons can give
rise to large numbers of isoforms. Thus, it has been
historically difficult for technology and statistical methods to allow
researchers to distinguish between different isoforms of the same gene.

%s1.2 ###
\subsection{Ultra High Throughput Sequencing}

Ultra High Throughput Sequencing (UHTS or simply ``sequencing'') is an
emerging technology
which promises to become as (or more) powerful, popular and
cost-effective than current microarray technology for several
applications, including isoform estimation.
When used to study mRNA
levels, UHTS is referred to as RNA-Seq. In the past
year, studies using UHTS to study genome organization, including
isoform expression, have been prominent (see \cite{Pan2008};
\cite{COX};
\cite{WAHL}; \cite{HANSEN}; \cite{MAHER}) and featured in the
journals Science and
Nature (see \cite{SULTAN}; \cite{Wang2008}), which dubbed 2007 as the
``year of
sequencing'' (see \cite{CHI}).

Briefly, UHTS is a method that relies on directly sequencing the
nucleotides in a sample rather than inferring abundance of mRNA by
measuring intensi\-ties
using predetermined homologous probes as micro\-arrays do. Thus, the data
generated from an UHTS
experiment are large numbers of discrete strings of nucleotides,
called base pairs (bp), which can take values of A, C, G or T.
In 2010, each
experiment produced tens of millions of up to 100bp reads. The throughput
of this technology is expected to continue its rapid growth.

Two experimental protocols for RNA-Seq are in
common use: (a) single end and (b) paired end sequencing
experiments. For single end experiments, one end (typically
about 50--100 bp) of a long (typically 200--400 nucleotide) molecule is
sequenced. For paired end experiments, typically 50--100 bp of both
ends of a typically 200--400 nucleotide molecule are sequenced.
Using current Illumina technology, each time the sequencing machine is
operated, eight samples (e.g., potentially eight different catalogues of
gene expression)
can be interrogated (essentially) independently and tens of millions of
reads are produced in each sample.

%s1.3 ###
\subsection{Related Work}

An important application and use of UHTS technology is to
quantify the abundance of mRNA in a cell (RNA-Seq). This is done by
matching the sequences generated in an UHTS experiment to a~data\-base of
known mRNA sequences (called alignment) and inferring the abundance of
each mRNA from the number of experimental reads (fragments of the original
mRNA molecules) aligning to it. Sometimes, a statistical model is used
for this
estimate. Importantly,
experimental steps involved in an UHTS experiment can affect the
probability of each fragment being observed, although modeling of
these processes is not the focus of this paper.

The rapid technological advances in sequencing have spawned a large
number of algorithms for analyzing
sequence data (see \cite{bowtie}; \cite{tophat}; \cite{cufflinks};
\cite{Mortazavi2008}), some of which aim
to estimate mRNA abundance.
%methodology for
%analyzing the results of isoform specific RNA-Seq experiments. Notable
%exceptions are the work of
%commonly generate data which require more
%complex statistical models than previously developed. }
To date,
inference on the abundance of mRNA has been made by aligning
reads to known genes and estimating a gene's expression by
averaging the number of reads which map uniquely to it using the
simplifying assumption that the transcript is sampled uniformly (see
\cite{Jiang2009};
\cite{Mortazavi2008}), and sometimes using heuristic approaches to
accommodate reads which map to multiple
locations (see \cite{Mortazavi2008}). These models do not provide optimal
estimators of isoform-specific expression levels and do not accommodate
modeling of
important steps in the experimental procedure. The work in this paper
significantly extends a basic
Poisson model developed in~\citet{Jiang2009} to allow for more flexible
and efficient inference and establish rigorous statistical theory. In
particular, the model in \citet{Jiang2009} does not work with paired
end sequencing data, or read-specific sample rate in a sequencing
protocol.

%efficiency between single and paired end
%experimental designs for sequencing experiments through a study of
%Fisher information.}

%isoform specific expression with larger Fisher information than
%previously
%proposed estimators. Further, this paper introduces a flexible model
%which
%can accommodate
%models of non-uniform sampling of a particular location in a
%transcript. Work presented here also advances statistical modeling and
%analysis of current methods
%for UHTS by providing a method for inference
%in expression studies when paired end reads are used. In addition, it
%quantifies the advantage of using paired end sequencing
%technology compared to single end technology for RNA-Seq. Statistical
%properties of the model such as minimal sufficiency are investigated.
%Using Fisher information and simulation, this paper establishes gains
%in accuracy of
%estimates of gene expression using paired end libraries. To our
%knowledge, this is the first such statistically rigorous methodology
%and analysis to be
%developed. }

This paper introduces a statistical model for estimating isoform
abundance from RNA-Seq data.
By grouping the reads into categories and modeling the read counts
within each category as Poisson
variables, the model is flexible enough to accommodate both single end
and paired end RNA-Seq data.
Based on the derivation of minimal sufficient statistics, a
computationally feasible implementation
of the maximum likelihood estimator of the model is provided. Using a
study of the Fisher information
and also numerical simulation, it is shown that using paired end
RNA-Seq one can get more accurate isoform
abundance estimates. To the best of our knowledge, this is the first
such statistically rigorous methodology and
analysis to be developed.

%s2 ###
\section{RNA-Seq}

Isoforms of a gene are subtle differences in a gene sequence, sometimes
resulting from inclusion or exclusion of a single exon, a discrete piece
of sequence depicted in Figure~\ref{fig:alternative_splicing}. In
principle, compared to microarrays, UHTS has the potential to
provide high resolution estimates of isoform use. However, signal deconvolution
must take place for these estimates to
be accurate.

In order to estimate the expression of different
isoforms of the same gene, several measurements of that gene's
expression, whether from a microarray or sequencing, must be
deconvolved.
Several studies have investigated this deconvolution problem when
measurements are made from a microarray (see \cite{HILLER} or
\cite{SHE}). This paper presents an estimator for deconvolution
for ultra high throughput sequencing experiments.

%To understand
%the data format, some background on UHTS experiments is needed.

\begin{figure}

\includegraphics{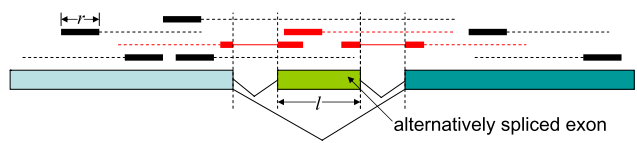}

\caption{Single end sequencing. A gene of three exons is shown with
the middle exon of length $l$ being
alternatively spliced. Reads that come from this gene are shown above
the gene in solid bars and the parts
that are not sequenced are shown in broken lines. Reads that span an
exon--exon junction are shown in solid
bars connected by thin lines. Reads that are related to the AS exon are
shown in red color. In this case only
the reads in red are isoform informative.}\label{fig:longer_reads}
\end{figure}

\begin{figure}[b]

\includegraphics{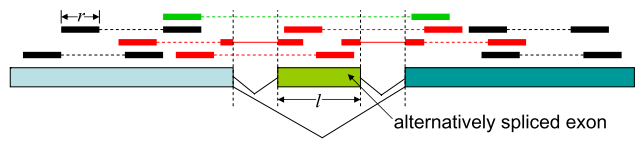}

\caption{Paired end sequencing. A gene of three exons is shown with
the middle exon of length $l$ being
alternatively spliced. Paired end reads that come from this gene are
shown above the gene in solid bars and
the parts that are not sequenced are shown in broken lines. Reads that
span an exon--exon junction are shown
in solid bars connected by thin line. Reads that are directly related
to the AS exon are in red as before.
Reads that provide indirect information for separating isoform
expressions are in
green.}\label{fig:paired_reads}
\end{figure}

As mentioned, two experimental approaches for RNA-Seq are in wide use.
In single end read experiments, reads are
generated from one end of a~molecule (depicted schematically in Figure
\ref{fig:longer_reads}); in paired end reads, reads are
generated from both ends of a mole\-cule, but typically a large number of
nucleotides interior to the molecule
are left unsequenced (depicted schematically in Figure \ref{fig:paired_reads}). The length
of the whole molecule being sequenced
is called the insert size or insert length.

%f2 ###

%f3 ###

To appreciate the additional information provided by the paired end
reads, consider
Figure~\ref{fig:longer_reads} which depicts single end reads randomly
sampled from a~transcript of a gene.
Suppose there are two possible isoforms for the transcript of this gene
depending on whether an exon of
length $l$ is retained or skipped. In this case, only the reads that
come from the alternatively spliced exon
(AS~exon), or come from junctions involving either the AS exon or the
two neighboring exons, can provide
information to distinguish the two isoforms from each other, that is,
only these reads are isoform informative.
If the AS exon is short compared to the transcript, then the majority
of the single end reads contain
information only on gene level expression but not isoform level
expression. Assuming uniform distribution on
the reads' positions in the gene, it is evident that a read is related
to the AS exon with probability
$P={\frac{l+r}{L-r}}$ if the read comes from the AS exon
inclusive isoform, where $L$ is the
length of the whole gene (without the intronic regions) and $r$ is the
length of the reads. Thus, $P$ is a
strictly increasing function with respect to the read length $r$ as
well as the AS exon length $l$. As an
example, for a gene of length 2000 bp with a short AS exon of length 50
bp, $P=0.0406$ for reads of length
30 bp, $P=0.0513$ for reads of length 50 bp, and $P=0.0789$ for reads
of length 100 bp.

Currently, technical limitations limit the length of sequenced reads.
These limitations vary by particular
platform used for UHTS. The two platforms in widest use are the
Illumina platform and the ABI SOLiD
platform. To date, the longest read that can be sequenced on the
Illumina platform is roughly 100 bp, and the
most reliable read length is still roughly 70 bp.\setcounter
{footnote}{1}\footnote{The read
length is roughly the same for the ABI SOLiD
platform. For the 454 platform the read length can be several folds
higher, but the throughput is much
lower compared to the other two platforms. Because sequencing
technology is
developing so rapidly, these numbers are likely to be out of date very
soon. Our statistical models apply
to all platforms and all read lengths.} %In the future, even if
%technological improvements
%enable longer high quality read lengths, experimental interest
%may still focus on relatively short fragments. Applications
%include ribosome footprinting
%footprinting \cite{YEO} which
%profile the interactions between RNA or DNA and proteins.%\jamie{why
%put this in last sentence in when this paper focuses on the benefit of
%paired end reads for isoform changes? These two examples wouldn't
%benefit from the analysis presented here. As with everything, just an
%opinion}

Paired end reads are an attractive way to decouple the isoform specific
gene expression. By performing paired
end sequencing, reads are produced from both ends of the fragments, but
the interior of the fragment remains
unsequenced. This method of sequencing both sides of the fragment
increases the number of isoform-informative
reads as illustrated in Figure \ref{fig:paired_reads}. Paired end
reads that are mapped to the genes are
shown in solid bars above the gene, with read pairs connected by broken lines.

As shown in Figure \ref{fig:paired_reads}, some read pairs (colored
red) are directly informative on the
retention or skipping of the AS exon. In addition, some read pairs span
both sides of the AS exon (colored
green). For these read pairs, the length of the fragment that they span
(a.k.a. the insert size or insert
length) depends on whether the AS exon is used or skipped in the
transcript. If the distribution of the
insert size is given, then these read pairs can also provide
discriminatory information on the isoforms as
shown in Figure \ref{fig:paired_reads} and developed rigorously
through the insert length model in Section
\ref{sec:insert_length_model}. For illustration, suppose the
experimental protocol selects fragments of sizes
around 200 bp for pair-end sequencing.\footnote{The insert size can be
controlled by tuning the parameters
involved in the fragmentation, random priming and size selection steps
in the sample preparation process.}
In such an experiment, if the insert size of a read pair is either
200 bp or 350 bp depending on whether the
read pair came from a transcript that included or excluded an exon of
length 150 bp, then this read pair is
unlikely to have come from a transcript that retained the AS exon.

It is easy to see from Figure~\ref{fig:paired_reads} that the fraction
of reads that contain information to
distinguish the two isoforms from each other increases not only with
the read length and the length of the AS
exon, but also with the insert size (when the insert size distribution is
a point mass). Since it is possible to have a much longer insert size
than read
length,\footnote{Current technology allows a biochemical
modification of sequenced molecules (via a circularization step) that
can produce two short reads
from two physical locations on a molecule that may be separated by up to
several kilobases (using the ABI platform or a long-insert protocol from
Illumina), which is also called the mate-pair sequencing. Although
technologically it is different from the paired end sequencing,
the analysis is the same from a statistical point of view.}  a
considerable amount of information can be extracted
from the paired end reads for decoupling the isoform-specific
gene expression. This concept is developed precisely in the
following sections.

%s3 ###
\section{The Model}\label{model}

%s3.1 ###
\subsection{Notation}

The notation in Table~\ref{table:notations} is used to present
the statistical model.

%t1 ###
\begin{table*}%[!htb]
\caption{Notation}\label{table:notations}
\begin{tabular}{@{}ll@{}}% was cl
\\[3pt]
\hline
\textbf{Symbol} & \multicolumn{1}{c@{}}{\textbf{Meaning}} \\
\hline
$I$ & Total number of unique transcripts (nucleotide sequences) in the
sample. \\
$J$ & Total number of unique reads. \\
$\theta_i$ & The abundance of transcript type $i$, $i=1,\ldots,I$. \\
${\theta}$ & The isoform abundance vector $[\theta_1,\theta_2,\ldots
,\theta_I]$. \\
$s_j$ & Read type $j$, $j=1,\ldots,J$. \\
$n_{i,j}$ & The number of reads $s_j$ that are generated from
transcripts $i$. \\
$n_j$ & The number of read $s_j$ that are generated from all the
transcripts, \\ & that is, $n_j=\sum_{i=1}^I n_{i,j}$. \\
$a_{i,j}$ & Up to proportionality, the sampling rate of $n_{i,j}$,
that is, the rate that \\ & read $s_j$ is generated from each individual
transcript $i$. \\
${a_j}$ & The sampling rate vector $[a_{1,j},a_{2,j},\ldots,a_{I,j}]$
for read $s_j$. \\
${\theta} \cdot a_j$ & The sampling rate of $n_j$, that is, the rate that
read $s_j$ is generated \\ & from all the transcripts. \\
$A$ & The $I\times J$ matrix of the sampling rates $\{a_{i,j}\}
_{i=1,j=1}^{I,J}$. \\
$c_i$ & The number of copies of the $i$th transcript in the sample.\\
$l_i$ & The length of the $i$th transcript in the sample.\\
$n$ & The total number of reads.\\
\hline
\end{tabular}
\end{table*}

%s3.2 ###
\subsection{Assumptions}
\label{modelassumptions}

The following assumptions on the process of
UHTS are used in this paper. %Such
%experiments generate large sample sizes (millions of reads
%generated in one experiment).

\begin{longlist}[(5)]
\item[(1)] The sample contains $I$ unique transcripts. In this paper we deal
with one gene at a time and consider
all the isoforms of the genes of interest as the set of $I$
transcripts. The abundances for the transcripts
are the parameters of interest and denoted $\{\theta_i\}_{i=1}^I$.
\item[(2)] After sequencing the sample, there are $J$ distinct reads denoted
as $\{s_j\}_{j=1}^J$. A type
of read refers to a single end read that is mapped to a specific
po-\vadjust{\eject} sition (which can be denoted as the $5'$ end
of the read) in a transcript in single end sequencing, or a~pair of
reads that are mapped to two specific
positions (which can be denoted as the $5'$ end of the first read and the
$3'$ end of the second read)
in paired end sequencing.  % This definition is made
%rigorous in Section $\ref{}$.
%
\item[(3)] Each transcript is
independently processed\break and then sequenced.
\item[(4)] $n_{i,j}$, the number of reads of type $s_j$ that are generated
from transcript $i$, are approximated as
Poisson random variables with parameter $\theta_ia_{i,j}$, whe\-re
$a_{i,j}$ is the relative rate that each individual transcript
$i$ generates read $s_j$, called the sampling rate defined below.
\item[(5)] Given $\{\theta_i\}_{1 \leq i \leq I}$, $\{n_{i,j}\}_{1 \leq i
\leq I, \; 1 \leq j \leq J}$ are independent
random variables.
\end{longlist}

If transcript $i$ cannot generate read $s_j$, $a_{i,j}$ is set to
zero: $a_{i,j}=0$. More specifically, for $1 \leq i_1, i_2 \leq I$, $1
\leq j_1,j_2 \leq
J$, assuming none of the $a_{i_k,j_k}$ for $k=1,2$ are zero,
$a_{i_k,j_k}$ are defined so that
\begin{eqnarray}\label{prob.ratio}
\frac{a_{i_1,j_1}}{a_{i_2,j_2}}&=&\operatorname{Pr}(\mbox{read
$s_{j_1}$ observed after}\hspace*{-18pt}\nonumber\\
&&\hphantom{\operatorname{Pr}(}\mbox{processing one copy of transcript
}i_1)\hspace*{-18pt}\nonumber\\ [-8pt]\\ [-8pt]
&&{}/\operatorname{Pr}(\mbox{read $s_{j_2}$ observed after}\hspace*{-18pt}\nonumber
\\
&&\hphantom{{}/\operatorname{Pr}(}\mbox{processing one copy of transcript }i_2).\hspace*{-18pt}\nonumber
\end{eqnarray}

Therefore, up to a
multiplicative constant, $a_{i,j}$ is the sampling rate of the
$j$th read from the $i$th transcript. This constant is chosen so
that the
estimates of $\theta_i$ are normalized in order to be comparable
across experiments. Two such choices are
described in Section~\ref{sec:sampling_rates}. With appropriate choice
of $a_{i,j}$, the probabilistic
interpretation of $a_{i,j}$ can be maintained across different
experiments.\footnote{The implementation of
the model described in this paper ignores reads that align to multiple
genes (while of course not ignoring
reads that align to multiple isoforms). This detail does not impact the
significant number of genes which
contain no such reads that map to multiple genes, and a simple
adaptation of the model can accommodate reads mapping to multiple genes.}

\begin{example}
Suppose a gene has three exons and two isoforms, as shown in Figures \ref{fig:longer_reads}
and  \ref{fig:paired_reads}. Suppose the three exons
have lengths 200 bp, 100 bp and 200 bp. Suppose the read length is 50 bp
and single end reads are generated
from a transcript uniformly. There are totally 500 different reads. 302
of them are from regions shared by
the two isoforms, 149 of them are from isoform 1 only and 49 of them
are from isoform 2 only. In this case,
$I=2$, $J=450$ and the matrix $A$, up to a multiplicative constant, is
\[{\fontsize{10.5pt}{12.5pt}\selectfont{
A=\pmatrix{
1 & 1 & \cdots& 1 & 1 & 1 & \cdots& 1 & 0 & 0 & \cdots& 0\cr
1 & 1 & \cdots& 1 & 0 & 0 & \cdots& 0 & 1 & 1 & \cdots& 1
},
}}\]
where $A$ has $302$ columns of $1\choose1$, $149$ columns of
$1\choose0$ and $49$ columns of
$0\choose1$.
\end{example}

%Since in model $\ref{}$ the $\theta_i$'s are the only quantities to
%estimate and they are coupled with the $a_{i,j}$'s all the time,
%$\theta_i$ is always non-identifiable up to a constant factor, i.e.,
%$k\theta_i$ and $\{a_{i,j}/k\}^I_{i=1}$ will provide the same fit to
%our model as $\theta_i$ and $\{a_{i,j}\}^I_{i=1}$.

%Therefore, it is sufficient to estimate the relative ratio between the
%$a_{i,j}$'s and to normalize the $\theta_i$'s at the end. \da{I would
%say we should make this more precise -- what do we mean by at the end?
%summing over all $\theta$?}

%s3.3 ###
\subsection{Likelihood Function}

The challenge of estimating isoform abundance arises from the
fact that different isoforms of a gene can have common sequence
characteristics and, therefore, different isoforms may generate common
read types. Thus, the $n_{i,j}$'s cannot be directly
observed. Rather, the observed quantities are sequences that are
necessarily collapsed over the potentially multiple transcripts
generating them. The observed quantities in an RNA-Seq
experiment are therefore $n_j$, where
\[
n_j:=n_{.,j}=\sum_{i=1}^In_{i,j},
\]
denoted as $n_j$ for simplicity.

Since $\{n_{i,j}\}_{1 \leq i \leq I, \; 1 \leq j \leq J}$ are assumed to
be independent, and it is assumed that the number of reads of type
$s_j$ that are generated
from transcript $i$ follows a
Poisson distribution with parameter $\theta_ia_{i,j}$, $n_j$
follows a Poisson distribution with parameter
$\sum_{i=1}^I\theta_ia_{i,j}=\ajtheta$, where
${\theta}$ is the vector of isoform abundance $[\theta_1,
\theta_2, \ldots, \theta_I]$ and ${a_j}$ is the
vector of sampling rates $[a_{1,j}, a_{2,j}, \ldots, a_{I,j}]$ for
read $s_j$, in which there is a component for each isoform.

Under the assumption that each read is independently generated,
given $\ms{\theta}{1}{I}$, $\{n_j\}_{j=1}^J$\vspace*{1pt} are independent Poisson
random variables, and therefore ha\-ve the joint probability
density function
\begin{equation}
\label{joint_likelihood}
f_{{\theta}}(n_1,n_2,\ldots,n_J)=\prod_{j=1}^J\frac{(\ajtheta
)^{n_j}e^{-\ajtheta}}{n_j!}.
\end{equation}

Note that since $\E(n_j)=\ajtheta=\sum_{i=1}^I\theta_ia_{i,j}$, for all
$i,j$, ${\theta}$, the density~(\ref{joint_likelihood}) is a
curved exponential family: the natural parameter of the model is
in $\mathbb{R}^J$ while the underlying parameter is in
$\mathbb{R}^I$ with $J>I$.

%s3.4 ###
\subsection{Statistical Models for the Sampling Rate:~$a_{i,j}$}
\label{sec:sampling_rates}

This paper focuses on two choices of $a_{i,j}$ and
illustrates the assumptions and interpretation of the resulting
$\ms{\theta}{1}{I}$ parameters. The two choices give
rise to two different models: the first is the
uniform sampling model, and the second is the insert
length model.

While these models differ by whether insert length is taken
into consideration, both are motivated by the same model of sample
preparation below. To facilitate such modeling, the biochemical steps
preparing a sample for sequencing are represented
schematically as the composition of the following:
\begin{longlist}[3.]
\item[1.] Transcript fragmentation: each full length\break mRNA is fragmented
at positions according to a~Poisson process with rate parameter
$\lambda$.\footnote{Because genomic coordinates are discrete, the occurrence times
in the Poisson process should be rounded to the nearest natural numbers.}
\item[2.] Size selection: each fragment is selected with some probability
depending on only its length.
\item[3.] Sequence specific amplification or selection:\break each sequence is
amplified or further selected based on sequence characteristics.
\end{longlist}

The sampling rate matrices $A$ for the uniform sampling model and
the insert length model presented below are approximated from the
same statistical model for steps (a) and (b) above. Namely,
transcript fragmentation (positions where the transcript is cut)
is modeled as a Poisson point process. Let $p(\cdot)$ denote the
probability mass function of fragment lengths obtained from this
process. Note that $p(\cdot)$ is an unobserved quantity because the
sample is subject to a size selection step after fragmentation and
before sequencing. The size selection step
is modeled as follows: a length $l$ fragment of transcript is
obtained with probability $r(l)$ independently of the identity of
the molecule. $r(\cdot)$ is called
the filtering function.

While the model in steps (a) and (b) are realistic across
experiments, modeling step (c) is more involved and variable
across experiments. Modeling how the specific nucleotide sequences
affect the
probability of being amplified and selected for sequencing varies
significantly by experiment and is beyond the scope of this work.
However, it is important to emphasize that the model presented in
this section is flexible enough to account for estimation of the
effect of step (c). Moreover, the model can be adapted to
accommodate different model choices in any of steps (a), (b) or
(c). In the two models presented below, it is assumed that
sequence selection and amplification are uniform.

\begin{figure*}

\includegraphics{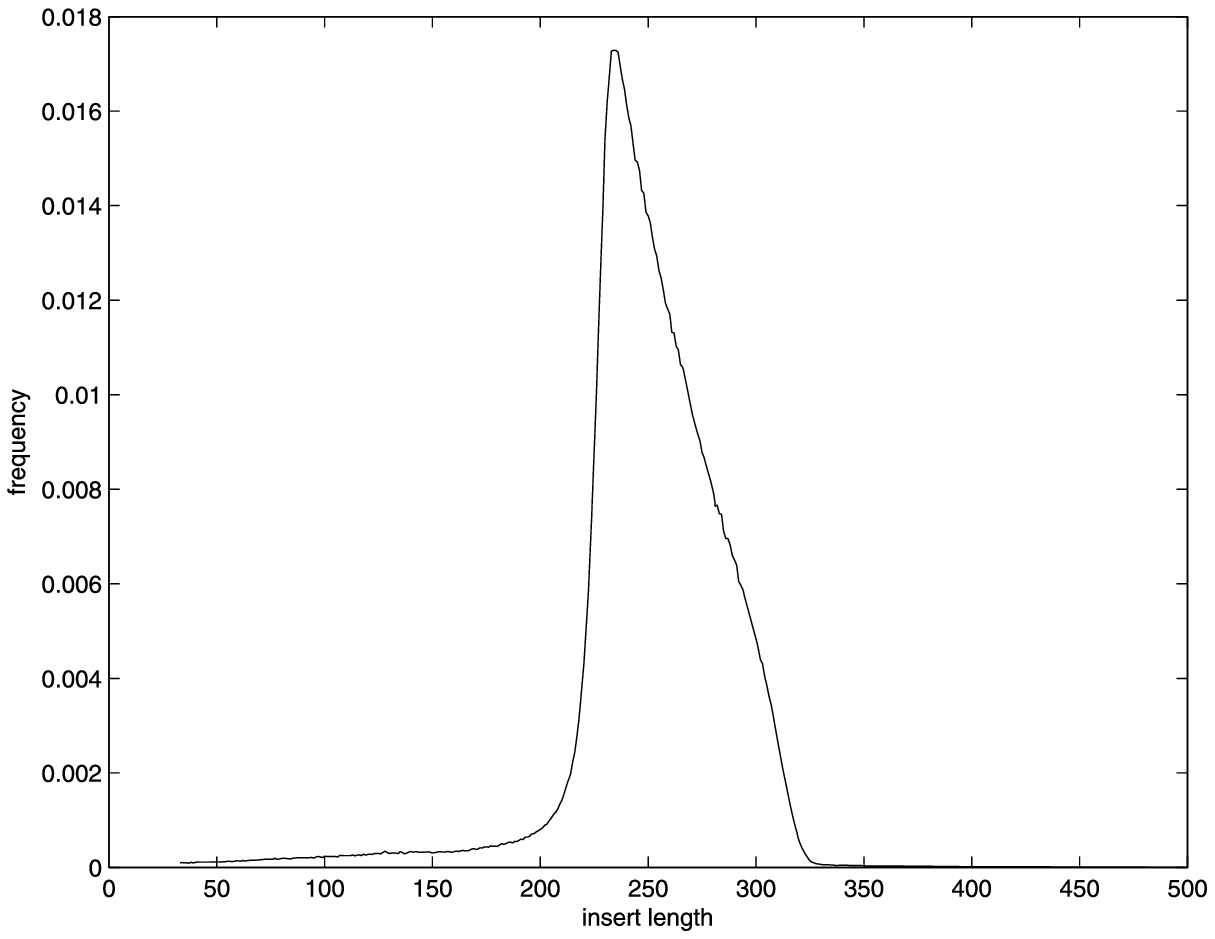}

\caption{A typical empirical mass function of the
insert length.}\label{fig:insert_length}
\vspace*{10pt}
\end{figure*}

Modeling the random processes (a) and (b) above as independent and only
dependent on a fragment's length and
assuming that sequence selection and amplification are uniform produces
a model for the distribution of
fragment lengths in the sample. This distribution is represented by
$q(\cdot)$ and can be estimated
empirically from a paired end sequencing run, namely, mapping both
pairs from each read to a database and
inferring the insert length.\footnote{In the traditional
bioinformatics literature this is also called
alignment, while the nomenclature ``mapping'' is more often used in the
UHTS literature where the sequences being aligned are
short reads.} Such an empirical function $\hat{q}(\cdot)$ is depicted
in Figure \ref{fig:insert_length} and
represents a reasonable approximation to the overall distribution of
molecule sizes sequenced in an
experiment. Further, note that a consequence of the modeling in steps
(a)--(c) above produces the
identity $q(l)=r(l)p(l)$.

Some mapping programs (such as introduced in \cite{bowtie}) have
options that take advantage of a user specified expected
insert size to help improve mapping performance, which may lead to
biases in the mapping. The mapping
procedure described in this manuscript performs each paired end
alignment by aligning the first and second
read separately, which does not bias the insert length~mo\-del and allows
for the calculation of minimal sufficient statistics for the model
and to perform statisti\-cal inference on isoform abundance without such~bias.

%f4 ###
\begin{figure*}
\begin{tabular}{@{}cc@{}}

\includegraphics{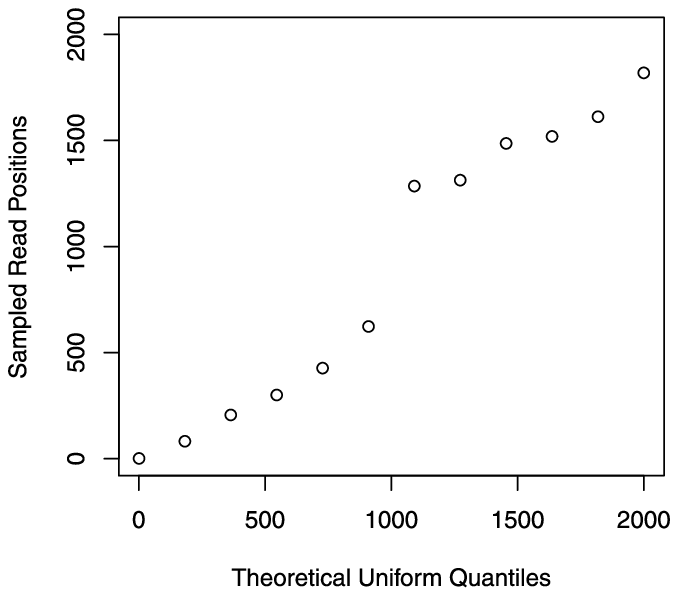}
&\includegraphics{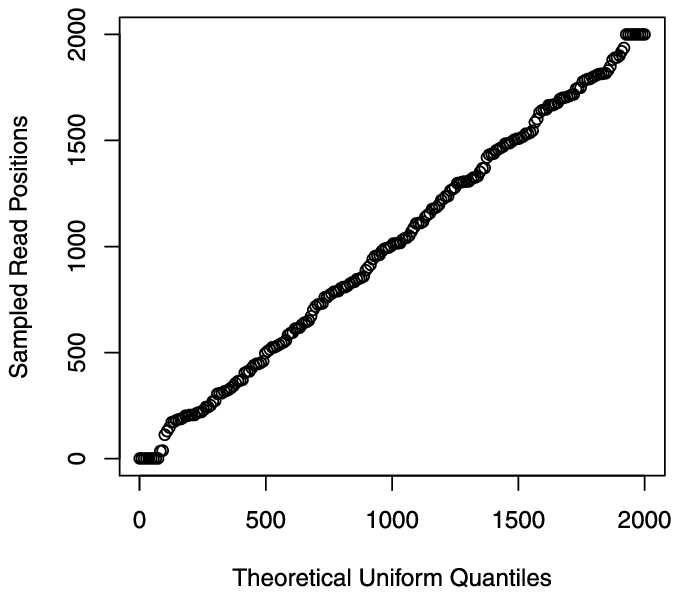}\\
(a)&(b)
\\
\multicolumn{2}{@{}c@{}}{
\includegraphics{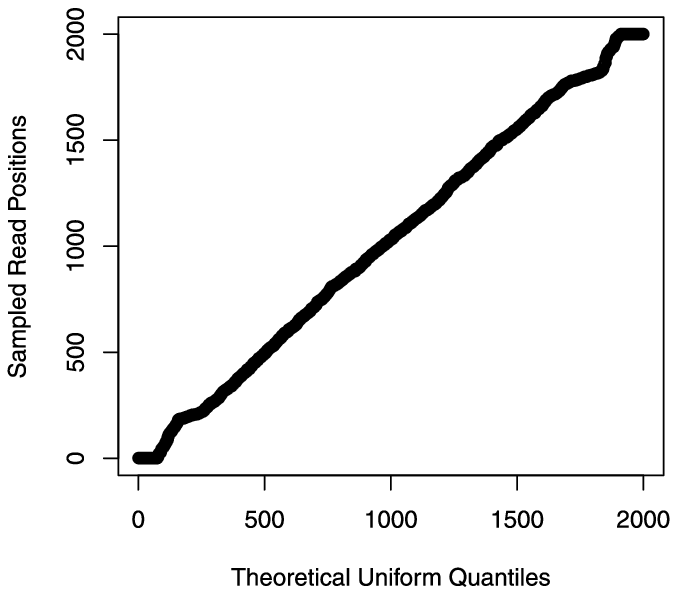}
}\\
\multicolumn{2}{@{}c@{}}{(c)}
\end{tabular}
\caption{Uniform Q--Q plot with sampled read
positions. \textup{(a)}, \textup{(b)} and \textup{(c)} are generated by simulations with 10,
100 and $1{,}000$ copies of transcripts,
respectively.}\label{fig:read_positions}
\vspace*{-4pt}
\end{figure*}

%s3.4.1 ###
\subsubsection{Uniform sampling model}

The uniform sampling model is
appropriate for single read data. It assumes
that during the sequencing process, each read (regarded as a
point) is sampled independently and uniformly from every possible
nucleotide in the biological sample. Uniform sampling is a good
approximation to sampling from a Poisson fragmentation process
and subsequent filtering step when the filtering function $r(\cdot)$
has support on a set that is small compared to the transcript
lengths; under these conditions, the process is approximately
stationary.

To investigate if the uniform sampling model satisfactorily
approximates the Poisson fragmentation and
filtering above for numerical regimes of transcript length and
fragmentation rate encountered in sequencing,
the following three simulations were performed: reads were generated
from 10, 100 or $1{,}000$ copies of a
transcript of length $2{,}000$ bp with $\lambda=0.005$. All the fragments
of length $200\pm20$ bp were retained and
the fragment ends were then compared to the sampled read positions as
modeled by the uniform sampling model
(see Figure~\ref{fig:read_positions}). It can be seen that as the
sample size increases, the two models are
very similar except at the two ends of the transcript. At the two ends
the Poisson process has some boundary
effects, and the sequencing protocol cannot be explained by a simple
model. For most situations, these
effects will be small, and hence are ignored in the uniform sampling model.

%%%%%%%%%%%%%%%%%%%%%%%%%%%%%%%%%%%%

%As shown
%in~\cite{Jiang2009}, this model works for short (unpaired) reads.

Thus, the uniform sampling model is appropriate for sequencing single
short reads where the sequencing process can be regarded as a
simple random sampling process, during which each read (regarded
as a point) is sampled independently and uniformly from every
possible nucleotide in the sample. The assumption of uniformity
implies that a constant sampling rate for all $a_{i,j}>0$ is
used. Specifically, let $a_{i,j}=0$ if transcript $i$ cannot
generate read $s_j$, and otherwise, $a_{i,j}=n$, where $n$ is the
total number of reads. As seen below, $n$ serves as a
normalization factor.

%%%%%%%%%%%%%%%%

To motivate this choice of $a_{i,j}$, consider the interpretation of
$\ms{\theta}{1}{I}$ induced by $A$. Under the uniform model, the
(unobserved) counts from the
$j$th nucleotide which is generated from the $i$th
transcript are modeled as a Poisson random variable with paramete $a_{i,j}
\theta_i$, that is,
\[
n_{i,j} = \operatorname{Po}(a_{i,j} \theta_i).
\]

%f5 ###

Computing $\E(n_{i,j})$ using the uniform sampling mo\-del with
$n$ total reads,
\begin{eqnarray*}
\E{(n_{i,j})}&=&n \operatorname{Pr}(\mbox{$j$th nucleotide generated by
transcript $i$})\\
&=& n \frac{ c_i }{ \sum_i l_i c_i},
\end{eqnarray*}
where $l_i$ is the length of the $i$th transcript and $c_i$ is
the number of copies of the $i$th transcript in the sample. Thus,
setting $a_{i,j}= n$ iff transcript $i$ can generate read
$j$ produces the identity
\[
n \theta_i = \frac{n c_i}{\sum_i c_i
l_i}
\]
so
%$$a_{C_f}=\sum_{i \in C_j} \alpha_i =l_f n$$ and hence with this
%choice of $\alpah_{i,j}$,
the uniform sampling model has parameter
\[
\theta_i= \frac{c_i }{
\sum_i l_i c_i}.
\]
This choice of $A$ has the property that it
normalizes $\ms{\theta}{1}{I}$ so that
\[
\sum_i \theta_i l_i=1,
\]
that is, it normalizes $\theta_i$ as a fraction of the total
nucleotides sequenced, as shown in~\citet{Jiang2009}, making it
conceptually compatible with the RPKM (Reads Per Kilobase of exon
model per Million mapped reads) normalization scheme
in~\citet{Mortazavi2008}, which is widely used by the RNA-Seq
community. This normalization conven-\vadjust{\eject}tion assumes the number of
nucleotides in the sequenced RNA of each cell does not vary between
samples. Modifying these assumptions to be more realistic yields better
choices for normalizing constants (see, e.g., \cite{normalizing}) and
can easily be
incorporated into the normalization of the sampling rate vector.

%By computing confidence intervals for the $\theta_i$ as outlined in
%different genes within a sample and of genes across samples. More
%thorough methodology for such procedures is the subject of current
%work.

%s3.4.2 ###
\subsubsection{Insert length model}
\label{sec:insert_length_model}

This model is applicable to paired end sequencing data. In paired end
sequencing, the insert length is usually controlled
to have a small range. Therefore, as suggested in
Figure~\ref{fig:paired_reads}, besides read positions,
information can also be extracted from insert lengths inferred
from reads. By modeling insert lengths properly, this piece of
information can be utilized and statistical inference can be
improved. Example~\ref{exp:insert_length} below illustrates this
concept and Section~\ref{sec:information} quantifies the gain in
statistical efficiency using the pairing information.

%Many factors such as
%fragment length, RNA structure and RNA sequence effect the sampling
%rate of particular sequences in an ultra high throughput experiment.
The insert length model models the sampling of transcripts,
conditional on insert length, as uniform. The insert length model
sets each $a_{i,j}$ using the empirical distribution of the insert
lengths of the sample (see Figure~\ref{fig:insert_length}) such
that conditional on the insert length, reads are sampled from
transcripts uniformly. This is specified mathematically as
\begin{equation}
\label{ins.eq}
a_{i,j}=q(l_{i,j})n,
\end{equation}
where $l_{i,j}$ is the length of corresponding fragment of $s_j$ on the
$i$th transcript, $n$ is the total
number of read counts and $q(l)$ is the probability of a fragment of
length $l$ in the sample after the
filtering. In application, for the insert length model, $q(\cdot)$ is
taken as $\hat{q}(\cdot)$, the
empirical probability mass function computed from all the mapped read
pairs. A~typical mass function is
illustrated in Figure~\ref{fig:insert_length}. Although usually this
function is unimodal (as in this case),
which favors our isoform estimation approach, our approach is flexible
enough to allow other types of
functions, such as bimodal functions, etc.

%%%%%%%%%%%%%%%%%%%%%%%%%%%%%%%%%%%%%%

To see the relationship between this choice of sampling rate
matrix and a model where reads are subject to Poisson
fragmentation and length dependent filtering, suppose that paired
end read $s_j$ is mapped to transcript 1 at coordinates
$(x_1,y_1)$ and transcript 2 at coordinates $(x_2,y_2)$ and both
reads are in the forward direction. Then, assuming none of
$x_1,x_2,y_1,\break y_2$ is at the boundary of a transcript, under the
Poisson fragmentation model (a) and length dependent size
selection (b),
\begin{eqnarray*}
&&
\operatorname{Pr}(\mbox{read }s_{j}\mbox{ observed after}
\\
&&\hphantom{\operatorname{Pr}(}\mbox{processing one copy of transcript }{i_1})
\\
&&\qquad{}/
\Pr(\mbox{read }s_{j}\mbox{ observed after}
\\
&&\hphantom{\qquad{}/\operatorname{Pr}(}\mbox{processing one copy of transcript }{i_2})
\\
&&\quad=
\operatorname{Pr}(\mbox{cut at $x_1, y_1$, no intermediate cut,}
\\
&&\hphantom{\quad=\operatorname{Pr}(}
\mbox{and transcript of length $x_1-y_1$ retained)}
\\
&&\qquad{}/
\operatorname{Pr}(\mbox{cut at $x_2, y_2$, no intermediate cut,}
\\
&&\hphantom{\qquad{}/
\operatorname{Pr}(}\mbox{and transcript of length $x_2-y_2$ retained)}
\\
&&\quad=
\operatorname{Pr}(\mbox{tr. of length $x_1-y_1$ retained $\vert$}
\\
&&\hphantom{\quad=
\operatorname{Pr}(\mbox{tr. of ,$\vert$}}\mbox{cut at $x_1, y_1$, no int. cut})
\\
&&\qquad{}\cdot\operatorname{Pr}(\mbox{cut at $x_1, y_1$, no int. cut})
\\
&&\quad\qquad{}/
\operatorname{Pr}(\mbox{tr. of length $x_2-y_2$ retained $\vert$}
\\
&&\hphantom{\mbox{ta i .ned $\vert$}\qquad{}/
\operatorname{Pr}(}\mbox{cut at $x_2, y_2$, no int. cut})
\\
&&\qquad{}\cdot\operatorname{Pr}(\mbox{cut at $x_2, y_2$, no int. cut})
\\[1pt]
&&\quad=
\frac{r(|x_1-y_1|)}{r(|x_2-y_2|)}\frac{p(|x_1-y_1|)}{p(|x_2-y_2|)}
\\[1pt]
&&\quad=
\frac{q(|x_1-y_1|)}{q(|x_2-y_2|)}.
\end{eqnarray*}

%As long as
%%
%$x_1, y_1$, no
%int. cut})}{\Pr(\text{cut at $x_2, y_2$, no int. cut})}
%%
%is approximately $1$ (e.g. if the expected length of a fragment is large
%compared to the difference between $x_1-y_1$ and $x_2-y_2$), }

Thus, the ratio
\[
\frac{a_{i_1,j}}{a_{i_2,j}}
\]
is approximately the same as defined by the sampling rate matrix
$A$ for the insert length model, with the assumption that none of
$x_1, x_2, y_1$ or $y_2$ is on the boundary of the transcript. As long
as the insert
length distribution has support which is small compared to
transcript length, relatively few transcripts map exactly to the
boundary, and little data is lost by ignoring them; doing so allows the
above conditions to be satisfied. Further, the argument above shows
that the
insert length model is consistent with assumptions (a)--(c) of
the sample preparation.

The insert length model yields a similar interpretation for the
normalization of $\ms{\theta}{1}{I}$ as in the uniform sampling
model, illustrated in the following computation: The paired end read
model specifies that the reads of type $j$ from transcript $i$
are Poisson with parameter
\[
n_{i,j} = \operatorname{Po}(a_{i,j} \theta_i).
\]

The insert length model assumes that reads are
filtered based on length independent of their sequence. This
produces a method of estimating the expectation of $n_{i,j}$. The
following approximates $\E(n_{i,j})$ under the insert length model:
\begin{eqnarray*}
\E{(n_{i,j})}&=& n \operatorname{Pr}(\mbox{read $j$ observed after}
\\
&&\hphantom{n \operatorname{Pr}(}\mbox{processing one copy transcript }i)\\
&:=& n\, \p{A \cap B \cap C},
\end{eqnarray*}
where $A$, $B$ and $C$ are defined as follows. Let $Y$ be a
random variable representing a read in the sample after
fragmentation. Let $A$ be the event that $Y$ is a~fragment of
transcript $i$, $B$ the event that $Y$ is read $j$ of transcript
$i$ and $C$ the event that $Y$ is a~fragment of length $l_{i,j}$
and is observed after filtering. Using the product rule,
\[
\p{A \cap B \cap C }= \p{B| A \cap C }\p{C|A }\p{A}.
\]

Each term is analyzed separately. Assuming uniform fragmentation across
the transcript and length dependent filtering,
\[
\p{B| A \cap C }= \frac{1}{l_i-l_{i,j}}.
\]

The basic assumption of the insert length model is that the
probability of observing a transcript of length $l_{i,j}$ does not
depend on the transcript and is equal to the empirical
insert length, $q(l_{i,j})$, hence % This assumption is valid if the
%size of each
% transcript is larger than $l_{i,j}$ and boundary effects are ignored.
%In this case,
%is close to 1. This assumption is justified by considering the Poisson
%fragmentation model where the rate of cutting $\lambda$ produces a mean
%insert length that is close to $l_{i,j}$ (see Weiss 1955). If this
%assumption is violated, by choosing $q(\cdot)$ to have support near
%$\lambda$,
%the approximations are still valid.
%}
\[
\p{C|A }\stackrel{\cdot}{=} q(l_{i,j}).
\]

To estimate $\p{\mathrm{A}}$, consider the random variables $X_i$, the
number of fragments in the sample from transcript $i$, and $X$, the total
number of transcript fragments in the sample. Then, assuming
transcript~$i$ is sufficiently and not overly abundant in the sample,\looseness=-1
\[
\p{A}= \E\biggl(\frac{X_i}{X}\biggr)\stackrel{\cdot}{=}\frac{\E(X_i)}{\E(X)}.
\]

Assuming a Poisson
fragmentation model, up to a boundary effect which has small impact on the
approximation,
\[
\frac{\E(X_i)}{\E(X)}\stackrel{\cdot}{=}\frac{c_il_i}{\sum_i
c_i l_i}.
\]

Combining these approximations yields
\[
\E{(n_{i,j})}\stackrel{\cdot}{=} n q(l_{i,j})\frac
{1}{l_i-l_{i,j}}\frac{c_il_i}{\sum_i c_i l_i}.
\]

Thus, if $\frac{l_i}{l_i-l_{i,j}}$ is close to $1$, $\theta_i$ is
identified in this model as
\[
\theta_i\stackrel{\cdot}{=}\frac{c_i}{\sum_i c_i l_i}.
\]

Thus, in both models, the choice of $a_{i,j}$ is consistent with its
definition in equation~(\ref{prob.ratio}). % Further, if instead
%of $l_f$ and $c_i$ denoting length and copy number of genes, they
%denoted length and copy number of categories, the computations
%above would hold for computing the expectation of $n_{C}$, the
%number of reads in a category. The normalization factor would be
%unchanged. \hui{category is not yet defined}
To illustrate the difference between the insert length and uniform
sampling models, consider the following example:

\begin{example}
\label{exp:insert_length}
Consider a case of two isoforms labeled $1$ and $2$ with an alternative
included exon as in Figure
$\ref{fig:alternative_splicing}$. Suppose the middle exon $2$ has
length 50 for concreteness. Suppose pair
end read $s_{j}$ has an imputed length of $50$ when mapped to $2$ and
of $100$ when mapped to $1$, as will be
the case if one of the ends is in exon $1$ and one in exon $3$.
Suppose the empirical
insert length function is modeled as uniform $[60,140)$. Then, in the
uniform model, because $n$ total
reads have been sequenced and mapped,
\[
a_{1j}=a_{2j}=n,
\]
whereas in the insert length model,
\[
a_{1j}=\frac{n}{80}\quad \mbox{and}\quad a_{2j}=0.
\]

Note that although the denominator $80$ in $a_{1j}$ in the insert
length model seems arbitrary, because there
are $80$ different paired end reads that start at the same position as
$s_j$, having all of them in the model
gives consistent gene expression estimates as in the uniform model.
\end{example}

%s3.5 ###
\subsection{Maximum Likelihood Estimation}
\label{sec:MLE}

In this paper $\theta$ is estimated using the MLE. Standard
theory shows that the MLE of model~(\ref{joint_likelihood})
will be consistent provided the
parameters in the model are in the interior of the parameter space
(see Theorem 6.3.10 of \cite{TPE}). Computationally efficient
procedures are needed to solve for these estimates in practice.

The fact that the density (2) is
Poisson allows for a~simplification of the calculation of the MLE by
regarding the parameter estimation as a generalized linear model (GLM)
problem with
Poisson density and identity link function (see \cite{McCullagh1989})
with extra linear constraints that require all the parameters
$\{\theta_i\}_{i=1}^I$ to be nonnegative. The optimization
problem in matrix form is
\begin{eqnarray}\label{optimization}
&&\mbox{maximize} \quad{n}^T\log({A^T}{\theta})-\mbox{sum}({A^T}{\theta})\nonumber
\\ [-8pt]\\[-8pt]
&&\quad\mbox{s.t. }{\theta} \geq0, \nonumber
\end{eqnarray}
where ${n}$ is a $J\times1$ column\vspace*{1pt} vector for the observed read counts
$[n_1,n_2,\ldots,n_J]$, $A$ is a
$I\times J$ matrix\vspace*{1pt} for the sampling rates $\{a_{i,j}\}_{i=1,j=1}^{I,J}$
and ${\theta}$ is the $I\times1$
isoform abundance vector $[\theta_1,\theta_2,\ldots,\theta_I]$.
$\log(\cdot)$ takes logarithm over each
element of a vector and $\mbox{sum}(\cdot)$ takes summation over all
the elements of a vector.

As shown in~\citet{Jiang2009}, the log-like\-lihood function
\[
\log(\mathcal{L}({\theta}))=\log(f_{{\theta}}(n_1,n_2,\ldots,n_J))
\]
is always concave and, therefore, any linear constraint convex
optimization method can be used to solve this
nonnegative GLM problem.\footnote{In our experiments we used the PDCO
(Primal-Dual interior method for
Convex Objectives, \href{http://www.stanford.edu/group/SOL/software/pdco.html}{http://}
\href{http://www.stanford.edu/group/SOL/software/pdco.html}{www.stanford.edu/group/SOL/software/pdco.html})
package developed by M. A. Saunders at Stanford University.}

\vspace*{2pt}
%s4 ###
\section{Sufficiency and Minimal Sufficiency}
\vspace*{2pt}

Because $J$ is usually very large, it is extremely inefficient to work
with the statistics $\ms{n}{1}{J}$ in
(\ref{joint_likelihood}) directly: in single end sequencing of a human
or mouse cell, $J$ can exceed $2{,}000$
for a typical gene, and in paired end sequencing with variable insert
length, it can easily reach $100{,}000$.
For computational purposes, it is therefore necessary to use sufficient
statistics for the likelihood
function~(\ref{joint_likelihood}). Because these statistics have an
intuitive interpretation, they are
referred to as a collapsing. This section analyzes sufficiency and
minimal sufficiency in model
(\ref{joint_likelihood}) and its relation to collapsing.

%s4.1 ###
\subsection{Sufficient Statistics and Collapsing}

As will be shown below, sufficient statistics have a natural
interpretation as collapsing read counts. Proposition~\ref{suff} shows
that to group reads $j$ and $k$ into the same
category, it is sufficient that reads have the same normalized
sampling rate vector (i.e.,
\[
\frac{{a_j}}{\Vert{a_j}\Vert}=\frac{{a_k}}{\Vert{a_k}\Vert},
\]
where
$\Vert\cdot\Vert$ is the vector 2-norm).

Such grouping of reads will be called (maximal) collapsings:
reads with the same normalized sampling rate vector are grouped
together. Intuitively, a maximal collapsing reduces the number
of such groups to be as small as possible.

\begin{definition}
Let $C_k$ be a collection of $m_k$ reads so
\[
C_k=\{s_{j_1}, \ldots, s_{j_{m_k}}\}_{1 \leq j_1<j_2 < \cdots<
j_{m_k} \leq J}.
\]

A set $\mathcal{C}=\{C_k\}_{k=1}^K$ is called a collapsing, if for any
$C_k \in\mathcal{C}$ and any $s_{j_1}, s_{j_2}\in C_k$,
\[
{a_{j_1}}=c{a_{j_2}}
\]
for some positive
number $c$.

Furthermore,
if for any $k_1\neq k_2$ and any $s_{j_1}\in C_{k_1}, s_{j_2}\in
C_{k_2}$,
\[
{a_{j_1}}\neq c{a_{j_2}}
\]
for any positive
number $c$, then $\{C_k\}_{k=1}^K$ is~cal\-led a~maximal collapsing. In a
collapsing, each $C_k$ is called a category.
\end{definition}

As will be seen in Theorem \ref{min.suff}, the maximal collapsing
gives rise to a set of minimal sufficient statistics, making it
useful from a computational perspective. A~real data example of such a
collapsing is provided in Section~\ref{applications}. The collapsed read
counts also have a~standard statistical interpretation as the sum
of independent Poisson random variables. Suppose categories
\[
\{C_k\mid k=1,2,\ldots,K\} \quad\mbox{with } C_k\subseteq\{s_1,s_2,\ldots
,s_J\}
\]
are nonoverlapping,
that is, $C_{k_1}\cap C_{k_2}=\varnothing$ when $k_1\neq k_2$. Then,
assuming each $n_j$ follows a Poisson distribution with parameter
$\theta
\cdot a_j$, $n_{C_k}$, the number of observed reads that belong to
category $C_k$ (i.e., $n_{C_k}=\sum_{s_j\in C_k}n_j$)
follows a Poisson distribution with parameter
\[
a^{(k)} \cdot\theta,
\]
where $a^{(k)}=\sum_{j=1}^{m_k} a_j^{(k)}$ and for $1 \leq j \leq
m_k$, $a_j^{(k)}$ is the sampling rate
vector of the $j$th read in category~$k$.

%The latter parameter can be
%interpreted as the relative probability that the $i^{th}$ transcript
%produced a read in the $k^{th}$ category.

\begin{proposition}
$\!\!\!\!$The maximal collapsing is unique.
\end{proposition}

\begin{pf}
The relation satisfied by two types of reads in a
category in the maximal collapsing is an equivalence
relation. This makes the maximal collapsing a grouping of reads
into equivalence classes which are always uniquely determined. To show
a relation is an equivalence relation, it suffices to
show that the reflexivity, symmetry and transitivity hold.

\textit{Reflexivity}: For any $s_j$,
\[
{a_j}={a_j},
\]
that is, $s_j\sim s_j.$

\textit{Symmetry}: For any $s_j$ and $s_k$,
\[
{a_j}=c{a_k}\quad\Rightarrow\quad{a_k}=\frac{1}{c}{a_j},
\]
that is, $s_j\sim s_k\Rightarrow s_k\sim s_j$.

\textit{Transitivity}: For any $s_j, s_k$ and $s_l$,
\[
{a_j}=c_1{a_k}
\]
and
\[
{a_k}=c_2{a_l}\quad
\Rightarrow\quad{a_j}=c_1c_2{a_l},
\]
that is,
$s_j\sim s_k$ and
$
s_k\sim s_l\ \Rightarrow\ s_j\sim s_l.
$
\end{pf}

To illustrate how maximal collapsing can be derived from the
choice of $a_{i,j}$ in the uniform model to produce the maximal
collapsing, reads with the same normalized sampling
rate vector are grouped into one category. Because $a_{i,j}$ is either $0$
or $n$, two reads $s_{j_1}$ and $s_{j_2}$ will have the same
normalized sampling rate vector, that is,
${a_{j_1}}/\Vert{a_{j_1}}\Vert={a_{j_2}}/\Vert{a_{j_2}}\Vert$, if and only if
they can be generated by the same set of transcripts.

\begin{example}
Consider a continuation of the setup in Example
$\ref{exp:insert_length}$. Suppose a uniform sampling model and
suppose reads $s_1$ and $s_2$ can be generated by both
transcripts $1$ and $2$, whereas read $s_3$ can only be generated
by transcript $1$. Then
\[
{a_1}={a_2}=[n,n]
\]
and
\[
{a_3}=[n,0].
\]

Grouping $s_1$ and $s_2$ together produces the maximal collapsing
$\mathcal{C}=\{\{s_1,s_2\},\{s_3\}\}$, the first category
containing reads that can be produced by both transcripts and the
second category containing reads only generated by transcript
$1$.
\end{example}

%More generally, with the maximal collapsing, the categories obtained
%are $\{C_S | S\subseteq\{1,2,\ldots,I\}, \exists
%s_i \text{ s.t. } a_{i,j}>0\Leftrightarrow j\in S \}$. Here is
%an example.

%s4.1.1 ###
\subsubsection{Collapsing and sufficiency}
Analysis of the likelihood
function~(\ref{joint_likelihood}) shows that collapsing the reads
produces sufficient statistics and maximal collapsings are
equivalent to minimal sufficient statistics.

Recall that a statistic $T(X)$ is sufficient for the parameter
$\theta$ in a model with likelihood function $f_\theta(x)$ if
\[
f_\theta(X)=h(x)g_\theta(T(X)).
\]

It is clear that the observed count vector
${n}=[n_1,n_2,\ldots,n_J]$ is sufficient for ${\theta}$. The collapsed
read count vector is also sufficient for $\theta$, as detailed in the
next proposition:

\begin{proposition}\label{suff}
For any collapsing ${C}=[C_1,\break C_2,\ldots,C_K]$, the
observed read count vector
${n_C}=[n_{C_1}, n_{C_2},\ldots,n_{C_K}]$ is a
sufficient statistic for ${{\theta}}$.
\end{proposition}

\begin{pf}
From the definition of collapsing, consider the $k$th category
$C_k$ with the re-enumerated reads $\{s^{(k)}_j\}_{1 \leq k \leq
K, 1 \leq j \leq m_k}$, the reads in category $k$ are enumerated
\[
C_k=\bigl\{s^{(k)}_1, s^{(k)}_2,\ldots,s^{(k)}_{m_k}\bigr\}.
\]

Define $a^{(k)}_j$ to be the sampling rate vector for
$s_j^{(k)}$, $1 \leq j \leq m_k$. By definition, for all $1 \leq
j \leq m_k$, for some scalar $c^{(k)}_j>0$,
\[
a^{(k)}_j= c^{(k)}_j a^{(k)}_1.
\]

Therefore,
\[
{{\theta}}\cdot{a^{(k)}_j}=c^{(k)}_j{{\theta}}\cdot
{a^{(k)}_1}.
\]

Rearranging the product in the right-hand side of equation~(\ref
{joint_likelihood}) as a
product over each read by the category into which it falls, and denoting
the $i$th read $n_i$ and parameter $\theta\cdot a_i$ as $x_j^{(k)}$
with parameter $\theta\cdot a_j^{(k)}$ when it falls as the $j$th
enumerated read in the $k$th category,
\begin{eqnarray}\label{eqn:min_suff}
&&f_{{\theta}}(n_1,n_2,\ldots,n_J)\hspace*{-8pt}\nonumber
\\
&&\quad=
{\prod_{i=1}^J\frac{(\theta\cdot a_i)^{n_i}e^{-{\theta\cdot a_i}}}{n_i!}}\hspace*{-8pt}\nonumber
\\
&&\quad=
{\prod_{k=1}^K\prod_{j=1}^{m_k}\frac{(\theta\cdot
a^{(k)}_j)^{x^{(k)}_j}e^{-{\theta\cdot a^{(k)}_j}}}{x^{(k)}_j!}}\hspace*{-8pt}\nonumber
\\[-8pt]\\[-8pt]
&&\quad=
{\prod_{k=1}^K\prod_{j=1}^{m_k}\frac{(c^{(k)}_j
\theta\cdot a^{(k)}_1)^{x^{(k)}_j}e^{-\theta\cdot a^{(k)}_j}}{x^{(k)}_j!}}\hspace*{-8pt}\nonumber
\\
&&\quad=
{\prod_{k=1}^K{\bigl( \theta\cdot a^{(k)}_1\bigr)^{\sum_{j=1}^{m_k}x^{(k)}_j}e^{-\sum_{j=1}^{m_k}
{\theta\cdot a^{(k)}_j}}}\prod_{j=1}^{m_k}\frac{(c^{(k)}_j)^{x^{(k)}_j}}{x^{(k)}_j!}}\hspace*{-8pt}\nonumber
\\
&&\quad=
%&={\prod_{k=1}^K\frac{({\theta} \cdot
%a^{(k)})^{x_{C_k}}e^{-{\theta}\cdot{a^{(k)}}}}{(
h(n_1,n_2,\ldots,n_J)g_{\theta}(n_{C_1},n_{C_2},\ldots,n_{C_K}),\hspace*{-8pt}\nonumber
\end{eqnarray}
where, since $\ms{n}{1}{J}= \{x_j^{(k)}\}_{1 \leq j \leq m_k, \; 1
\leq k
\leq K}$,
\[
h(n_1,n_2,\ldots,n_J)={\prod_{k=1}^K{
{\prod_{j=1}^{m_k}\frac{(c^{(k)}_j)^{x^{(k)}_j}}{x^{(k)}_j!}}}}
\]
and
\[
g_{\theta}(n_{C_1},n_{C_2},\ldots,n_{C_K})={\prod
_{k=1}^K\bigl(\theta}\cdot
a^{(k)}_1\bigr)^{n_{C_k}}e^{-\theta\cdot a^{(k)}},
\]
establishing the sufficiency of
${n_C}=[n_{C_1},n_{C_2},\ldots,\break n_{C_K}]$.
\end{pf}

%When all $c_j^{(k)}$ are equal, the computation above is simplified
%and the joint density of the collapsed read counts is clearly Poisson.

In addition to the sufficiency proved in Proposition~\ref{suff},
${n_C}$ is minimal sufficient if the corresponding collapsing
${C}$ is a maximal collapsing. This is detailed in the next
section.

%s4.2 ###
\subsection{Minimal Sufficiency}

To prove that the read counts derived from a maximal collapsing are
minimal sufficient statistics, recall
the following:

\begin{definition}[(Definition 6.2.13 of (Casella and Berger, \citeyear{CASELLA})]
For the family of
densities $f_{\theta}(\cdot)$, the statistic $T(X)$ is minimal
sufficient if and only if
\[
\frac{f_{\theta}(x)}{f_{\theta}(y)} \mbox{ does not depend on
$\theta$}\quad
\Leftrightarrow \quad T(X)=T(Y)
\]
\end{definition}

\begin{theorem}
\label{min.suff}
In the likelihood specified by equation \textup{(\ref{joint_likelihood})},
counts on maximally collapsed categories are minimal
sufficient statistics.
\end{theorem}

\begin{pf}
Let $T(X)$ be the collapsed vector of\break counts
$x_{C_1},x_{C_2},\ldots,x_{C_K}$ and let $T(Y)$ be the vector of
counts $y_{C_1},y_{C_2},\ldots,y_{C_K}$, each of
which are maximal collapsings. If $T(X)=T(Y)$,
equation~(\ref{eqn:min_suff}) shows that the ratio of densities
\[
\frac{f_{\theta}(x)}{f_{\theta}(y)}
\]
does not depend on~$\theta$. To
show the reverse implication, suppose $T(X) \neq T(Y)$. To show that
\[
\frac{f_{\theta}(x)}{f_{\theta}(y)} \mbox{ must depend on~$\theta$},
\]
it suffices to show that
\[
\frac{g_{\theta}(x)}{g_{\theta}(y)} \mbox{ must depend on~$\theta$}.
\]

It is
possible to simplify this ratio as
\begin{equation}
\label{factor}
\frac{g_{\theta}(x)}{g_{\theta}(y)}= \frac{ \prod
_{g \in G}(\theta
\cdot a_g)^{n_g}}{\prod_{h \in H}(\theta
\cdot a_h)^{n_h}},
\end{equation}
where $\{n_g\}_{g \in G}$ and $\{n_h\}_{h \in H}$ are positive numbers
and $G$ and $H$ are subsets of the
categories and are disjoint since if $G$ and $H$ share a common $j$,
the ratio in equation (\ref{factor})
can be reduced. Further, since the collapsings are maximal, for any
$a_i, a_j$ appearing in any product in
the numerator or denominator, there is no $c$ so that $a_i=ca_j$. Using
these properties, it will be shown
that the ratio of densities must depend on~$\theta$ by contradiction.

Suppose for some (now fixed) $T(X) \neq T(Y)$,
equation~(\ref{factor}) does not depend on $\theta$ and is
equal to a constant~$c$. Note that since $\theta$ can be the
vector of all $1$'s, if equation~(\ref{factor}) does not depend on
$\theta$, $c>0$ as when $\theta$ is the vector of all 1's both the
numerator and denominator of equation~(\ref{factor}) are positive. Then
equation~(\ref{factor}) is equivalent to a polynomial equation
\begin{equation}
\label{poly.factor} 0= c \prod_{h \in H}(\theta
\cdot a_h)^{n_h}- \prod_{g \in G}(\theta
\cdot a_g)^{n_g}
\end{equation}
$\forall\theta\in(\mathbb{R}^+)^I$. By basic algebraic
geometry, any polynomial in $\theta$ which is identically zero in
the space $(\mathbb{R}^+)^I$ is identically zero in all of
$\mathbb{R}^I$. Therefore, the last step is to show that the
right-hand side of equation~(\ref{poly.factor}) is not actually
zero for some $\theta\in\mathbb{R}^I$. To proceed, fix $h \in
H$. The claim is that there exists $v \in\mathbb{R}^I$ with $
\langle v, a_h \rangle=0$ but $\forall g \in G$,
\[
\langle v, a_g \rangle\neq0.
\]

This $v$ will be the choice of $\theta$ producing the
contradiction. For a vector $z \in\mathbb{R}^I$, let
$z^{\perp}$ denote the $(I-1)$-dimensional subspace of vectors
orthogonal to it. Then, to finish the proof, it suffices to
showing that there is some vector in $a_h^{\perp}$ which is not
in $\bigcup_{g \in G} a_g^{\perp}$. It is equivalent to show there
is a strict containment
\[
\biggl(\bigcup_{g \in G} a_g^{\perp}\biggr) \cap a_h^{\perp} = \bigcup_{g \in G}
(a_g^{\perp} \cap a_h^{\perp}) \subset a_h^{\perp}.
\]

Strict containment follows since for any $h \in H$,
\[
a_h^{\perp} \cap a_g^{\perp}
\]
is a subspace of dimension at most
$I-2$, thus, a~countable union of such spaces cannot equal a subspace of
dimension $I-1$.
\end{pf}

%%
%Let $\theta\in\mathbb{R}^d$ and $p(\theta)$ be a polynomial in
%$\theta$. Suppose $\forall\theta\in(\mathbb{R}^+)^I$, $p(\theta)=0$.
%Then, $\forall\theta\in\mathbb{R}^I$, $p(\theta)=0$.
%
%The proof is by basic algebraic geometry.
%
%For $1 \leq i <d$, define the space $\mathbb{R}_{i,+}$ as a subset of
%$\mathbb{R}^d$ where the first $i$ coordinates are in $\mathbb{R}$
%and the
%last $d-i+1$ coordinates are in $\mathbb{R}^+$:
%
%
%The proof is by induction on $m$, showing that if for $0 \leq m \leq d-1$,
%$p(\theta)=0$ for all $\theta\in\mathbb{R}_{i,+}$ then $p(\theta
%)=0$ for
%all $\theta\in\mathbb{R}_{i+1,+}$.
%
%To prove the inductive step, for $i \neq m+1$, fix $\theta\in\mathbb
%{R}_{i,+}$ in the space for which
%$p(\theta)$ is assumed to be zero and view $p(\theta)$ as a
%polynomial in
%the variable $\theta_{m+1}$ denoted $p_{\hat{\theta}_{m+1}}(\theta
%_{m+1})$. Since
%$p$ is a polynomial, there is an open interval
%$(a,b)$ so that if $\theta_{m+1} \in(a,b)$, $p_{\hat{\theta
%}_{m+1}}(\theta_{m+1})=0$. Since a non-zero
%polynomial can only have finitely many zeros, this implies that
%$\forall
%$p_{\hat{\theta}_{m+1}}(\theta_{m+1})=0$. Hence,
%%
%%
%$p(\theta)=0$. This completes the inductive step as well as base case
%for the induction.
%
%%
%}}

Using Theorem~\ref{min.suff}, the optimization problem [equation~(\ref
{optimization})] is reduced to
\begin{eqnarray}\label{optimization.min}
&&\mbox{maximize} \quad {n}^T\log({A^T}{\theta})-\mbox{sum}({A^T}{\theta})\nonumber
\\[-8pt]\\[-8pt]
&&\quad\mbox{s.t. }  {\theta} \geq0,\nonumber
\end{eqnarray}
where ${n}$ is a $K\times1$ column vector for the collapsed read
counts for categories $C_1,C_2,\ldots,C_K$,
$A$ is a $I\times K$ matrix for the collapsed sampling rates and
${\theta}$ is the isoform abundance vector.

The next section illustrates the relationship of minimal
sufficient statistics to raw data observed in sequencing
experiments.

%s5 ###
\section{Application}
\label{applications}

This section illustrates how minimal sufficient sta\-tistics are
calculated in an example with real RNA-Seq
data from an experiment on cultured mouse B cells. After the sequencing
reads were generated, they were
mapped to a data\-base of known mouse mRNA transcripts using the RefSeq
annotation data\-base (see \cite{RefSeq})
and the mouse reference genome (mm9, NCBI Build 37). The reads were
mapped using SeqMap, a short sequence
mapping tool developed in \citet{Jiang2008}. The two ends of the paired
end reads were mapped separately and
then a filtering step was applied during which only the pair of reads
which were mapped to the same
transcript and on the right direction were retained. Further, in the
analysis of this section, reads that map
to multiple genes were also discarded for computational ease. Because
we are mapping the reads to transcript
sequences rather than the whole genome, the positions that cannot be
uniquely mapped are less than 1\%, which
is not likely to change our results significantly. Of course, the model
presented in Section \ref{model}
can accommodate reads which map to multiple genes because of the
statistical equivalence of this problem to
that of deconvolving the expression levels of multiple isoforms. We
have chosen not to implement this
approach because only a small number of genes are impacted and because
as rapid growth of the technology
continues to produce longer reads, the problem will become negligible.
A total of 2,789,546 read pairs (32 bp
for each end) passed the filtering. The empirical distribution of the
insert length was inferred. This
distribution has a mean of 251 bp and a single mode of 234 bp (See
Figure~\ref{fig:insert_length}).

%f6 ###
\begin{figure*}

\includegraphics{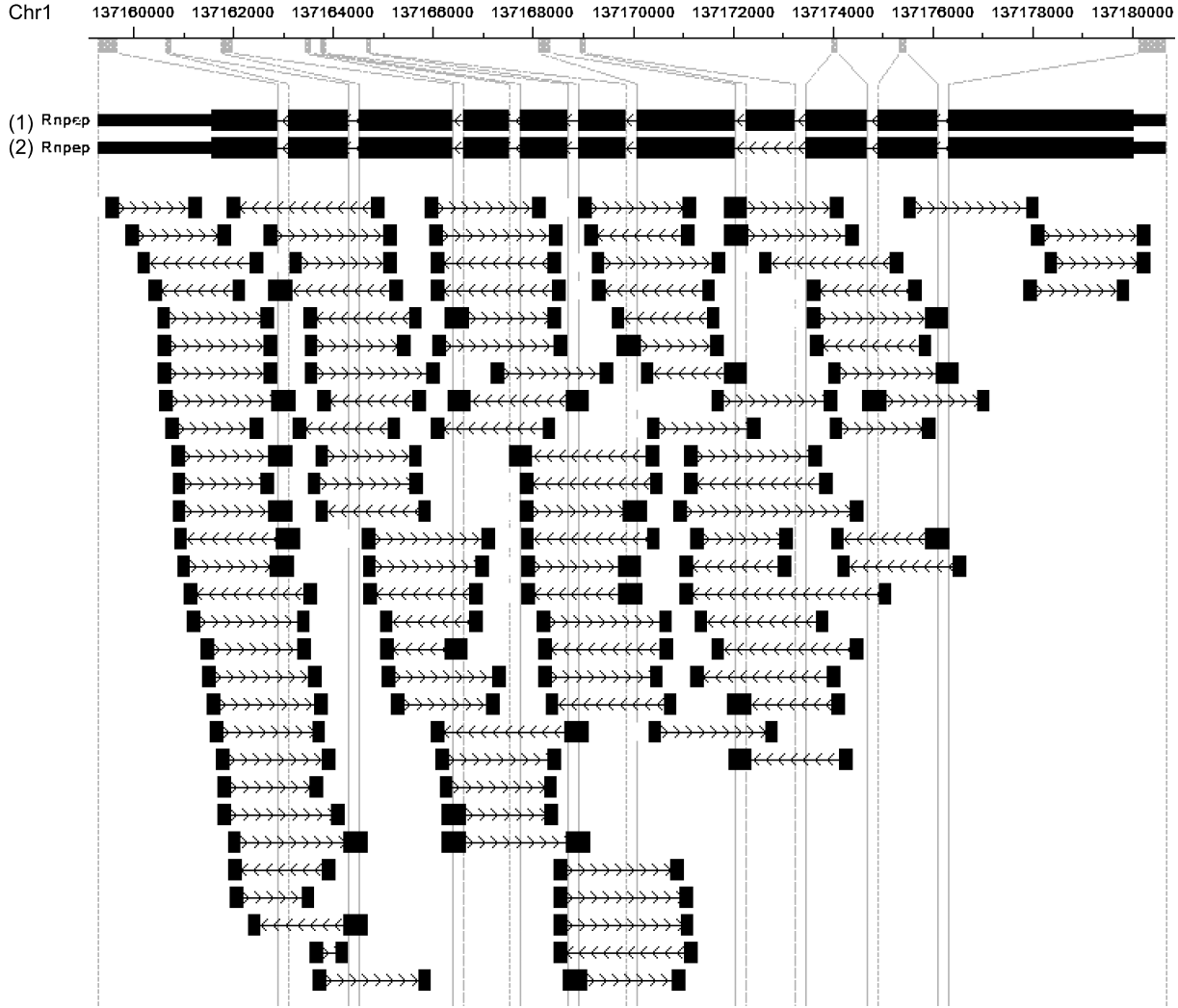}

\caption{Visualization of RNA-Seq read pairs mapped to the mouse gene
Rnpep in the CisGenome Browser (see
Jiang et al., \citeyear{Jiang2010}). From top to bottom: genomic coordinates, gene
structure where exons are magnified for
better visualization, read pairs mapped to the gene. Reads are 32 bp at
each end. A read that spans a
junction between two exons is represented by a wider box.}\label{fig:Rnpep}
\end{figure*}

Because more than $99\%$ of the data have an inferred length between
$73$ bp and $324$ bp, reads outside of
this range are not considered in subsequent analysis for this example,
as it is likely these reads come from
unannotated isoforms. This resulted in 27,118 (about $1\%$) read pairs
being excluded and the rest 2,762,428
(denoted as $n$ below) read pairs were used in the computation.\vadjust{\goodbreak}

The mouse gene Rnpep is used to demonstrate the computation of minimal
sufficient statistics. Rnpep has an alternatively spliced exon which
gives rise to two different isoforms (see
Figure~\ref{fig:Rnpep}). The gene itself is an amino peptidase, meaning
that it is used to degrade proteins in the cell. After mapping, 116
read pairs were
assigned to this gene, out of which 113 read pairs were used in
the computation after outlier removal. Figure~\ref{fig:Rnpep}
presents the positions where the reads are mapped. The gene was picked
because it has two alternatively spliced isoforms with a structure that
makes distinguishing reads from each isoform challenging, and because the
number of reads was small enough to visualize all of them in a
simple figure.

%t2 ###
\begin{table}[b]
\caption{Single end read categories for Rnpep}\label{tab:Rnpep_single_categories}
\begin{tabular}{@{}lcc@{}}
\hline
\textbf{Category ID} & \textbf{Sampling rate vector} & \textbf{Read count}\\
\hline
1 & $[4{,}242n, 4{,}242n]$ & 216\\
2 & $[296n, 0]$ & \phantom{0}10\\
3 & $[0, 62n]$ & \phantom{00}0\\
\hline
\end{tabular}
\end{table}

%s5.1 ###
\subsection{Uniform Sampling Model}

Any paired end read experiment can be treated as a single end read
experiment by taking each paired end read
and treating it as two distinct single end reads, one from each side of
the pair. In this, the 113 paired
end reads become 226 single end reads (without pairing information).

In the uniform sampling model, for either isoform, the sampling rate
vector for each read $s_j$ can take at most two values: $2n$ when the
isoform can generate read $j$ and $0$ when it cannot.
Because there are only two isoforms, one of which (isoform $2$)
excludes one of the exons
of the other (isoform $1$), it is evident that in the uniform sampling
model, there are only three
categories for the two isoforms.

The total length of isoform $1$ is $2{,}300$. The total length of
isoform $2$ is $2{,}183$. Hence, computing $a_{i,j}$ by summing over
the sampling rate vectors of the reads in the same category, the
three categories can be represented by their sampling vectors:
$[4{,}242n,4{,}242n]$, $[296n,0]$, $[0,62n]$. Using minimal sufficient
statistics reduces the data from a vector representing counts
on the $2{,}300$ possible reads $s_j$ from the two isoforms to the
$3$ minimal sufficient statistics which are counts on these
categories.

%
%To compute the minimal sufficient statistics in the uniform sampling
%model, note that the two isoforms of Rnpep produce just 3 categories.
%This reduces
%
%By enumerating all the positions
%we get totally 2,300 possible reads, some of them are tabulated
%in Table~\ref{tab:Rnpep_single_reads}. The third column of
%Table~\ref{tab:Rnpep_single_reads} gives the observed count for
%each particular read.

%%
%{tab:Rnpep_single_reads}} \centerline{
%%
%Category ID & Sampling rate vector & Read count\\
%1 & $[2n,2n]$ & 1\\
%2 & $[2n,2n]$ & 0\\
%3 & $[2n,0]$ & 0\\
%4 & $[2n,0]$ & 1\\
%5 & $[0,2n]$ & 0\\
%6 & $[2n,2n]$ & 1\\
%7 & $[2n,0]$ & 0\\
%8 & $[0,2n]$ & 0\\
%2300 & $[2n,2n]$ & 0\\
%%
%}
%%
%}
%
%}

The three categories representing minimal sufficient statistics are
tabulated in
Table~\ref{tab:Rnpep_single_categories}. Each category refers to a
group of reads that is generated by a
particular set of isoforms. For example, category 1 consists of reads
generated by both isoforms and category
3 consists of reads generated by isoform 2 only. Using these statistics
to solve the optimization
problem~(\ref{optimization}), the MLE for the two isoforms is $[\hat
{\theta}_1, \hat{\theta_2}] = [15.47,
2.70]$.\footnote{All the expression estimates in this paper are in
units compatible with RPKM (Reads Per
Kilobase of exon model per Million mapped reads) (see \cite
{Mortazavi2008}).} Bayesian credible intervals for
these estimates can be obtained by sampling from the posterior space of
the parameters (as outlined
in~\cite{Jiang2009}), the marginal $95\%$ credible intervals for
$\theta_1$ and $\theta_2$ are $(7.89,
18.81)$ and $(0.25, 10.83)$, respectively.

%s5.2 ###
\subsection{Insert Length Model}

To visualize how the insert length model can be used to produce
potentially stronger statistical inference as compared to the uniform
sampling model, consider Figure \ref{fig:Rnpep}. Each paired end
read is
depicted by two boxes with arrows joining pairs of reads. The direction
of the arrows represent which side of the read was sequenced first. For
those interested, the
direction of the arrows in the Rnpep gene itself indicates the
transcriptional direction of the gene in genomic coordinates, although
this concept can be ignored for the purposes here. Note that there is no
direct evidence that isoform $2$ is present in the sample, as no read
crosses the junction between the two exons which are adjacent in isoform
$2$ but not in isoform $1$. There is direct evidence of the presence of
isoform $1$, for example, as depicted in the fifth read from the left in
the first row which directly crosses a junction between two exons only
adjacent in isoform~$1$.

%t3 ###
\begin{table}[b]
\caption{Paired read categories for
Rnpep}\label{tab:Rnpep_paired_categories}
\begin{tabular}{@{}lcc@{}}
\hline
\textbf{Category ID} & \textbf{Sampling rate vector} & \textbf{Read count}\\
\hline
\phantom{00}1 & $[1{,}681.82n, 1{,}681.82n]$ & 95\\
\phantom{00}2 & $[294.60n, 0]$ & 10\\
\phantom{00}3 & $[0, 245.80n]$ & \phantom{0}2\\
\phantom{00}\vdots& \vdots& \phantom{0}\vdots\\
138 & $[0.0057n, 0.0018n]$ & \phantom{0}0\\
\hline
\end{tabular}
\end{table}

Because of the small gap between exons in the figure, reads spanning exons
will be slightly longer than reads not spanning exons. Also, some
inserts are
very short, and absence of the arrow connecting two reads indicates that
the entire insert has been fully sequenced. Note that several of the
reads spanning the alternatively spliced exon are exceedingly long.
This suggests that such reads are actually generated from
isoform $2$ rather than isoform $1$. If such reads are generated from
isoform $2$, they would likely have a smaller insert length than the
inferred insert
length when generated by isoform $1$, which are the lengths depicted in
the figure. Because the empirical insert length distribution has its
only mode near $250$ bp, conditional on observing the $6$th and
$7$th reads from the top of the figure spanning the alternatively
spliced exon, the read is more likely to come from isoform $2$. Thus,
there is indirect evidence of the presence of isoform $2$ in the sample.

Such indirect evidence is utilized by the insert length model; the model
produces quantitative estimates of the relative abundance of the two
isoforms. As will be seen in the next section, the abundance estimates
from the insert length model have larger Fisher information than the
estimates from the uniform sampling model.

In the insert length model, each of the possible insert lengths where
$q(\cdot)$ has support produce a unique read $s_j$ yielding
a total of $569{,}205$ possible
reads from the two isoforms. The maximal collapsing produces a total of
138 categories, some of which are represented in
Table~\ref{tab:Rnpep_paired_categories}. For intuition, all of the reads
with a fixed insert length
where both ends fall in the leftmost 7 or rightmost 3 exons of Rnpep
will be in the same category, as they have the same probability of
being sampled.

Using the minimal sufficient statistics, the MLE is computed to be
$[\hat{\theta}_1, \hat{\theta}_2] =
[16.73, 3.43]$. The marginal $95\%$ credible intervals for $\theta_1$
and $\theta_2$ are $(11.22, 21.02)$
and $(1.03, 9.29)$, respectively. The computed marginal $95\%$ credible
intervals for $\theta_1$ and
$\theta_2$ are nonoverlapping, whereas in the single end read model,
one cannot conclude that the expression
of isoforms $1$ and $2$ differ. Further, the insert length model has
slightly smaller marginal credible
intervals for each parameter.

This example suggests
that although the uniform sampling model for single end reads has twice
the sample size compared with the
insert length model for paired end reads, the insert length model
actually provides estimates with smaller
standard errors than those generated by the uniform sampling model,
because the insert length model can
utilize the extra information from the insert sizes of the reads. This
difference can be quantified by
analyzing the Fisher information of each model, the subject of
Section~\ref{sec:information}.

%s5.3 ###
\subsection{Practical Implementation Issues}

In general, to apply Theorem~\ref{min.suff}, one needs to enumerate
all the read types before collapsing, as
shown in the example of mouse gene Rnpep. This might be a time
consuming step, especially when the number of
read types is large. In practice, however, under some suitable sampling
rate models (which include both our
uniform model and insert length model), it is sufficient to enumerate
only the read types that have at least
one read being mapped. This can reduce the computation when the number
of mapped reads for the gene is small,
or, in other words, when the gene is lowly expressed.

To see how this works, rearrange the right-hand side of equation~(\ref{joint_likelihood}) as follows:
\begin{eqnarray}
&&
f_{{\theta}}(n_1,n_2,\ldots,n_J)\nonumber
\\
&&\quad=
{\prod_{j=1}^J\frac
{(\theta\cdot a_j)^{n_j}e^{-{\theta\cdot a_j}}}{n_j!}}\nonumber
\\
&&\quad=
{\prod_{n_j>0}\frac{(\theta\cdot a_j)^{n_j}}{n_j!}
\prod_{n_j=0}\frac{(\theta\cdot a_j)^{n_j}}{n_j!}
\prod_{j=1}^J e^{-{\theta\cdot a_j}}}\nonumber
\\ [-8pt]\\ [-8pt]
&&\quad=
{\prod_{n_j>0}\frac{(\theta\cdot a_j)^{n_j}}{n_j!}
\prod_{j=1}^J e^{-{\theta\cdot a_j}}}\nonumber
\\
&&\quad=
{\prod_{n_j>0}\frac{(\theta\cdot a_j)^{n_j}}{n_j!}
\prod_{j=1}^J{e^{-{\sum_{i=1}^I\theta_i{a_{i,j}}}}}}\nonumber
\\
&&\quad=
{\prod_{n_j>0}\frac{(\theta\cdot a_j)^{n_j}}{n_j!}
\prod_{i=1}^I{e^{-\theta_i\sum_{j=1}^J{a_{i,j}}}}},\nonumber
\end{eqnarray}
where only the term ${\sum_{j=1}^J{a_{i,j}}}$ depends on
the sampling rates of read types with
read counts $n_j=0$. Therefore, if we can compute this term
without\break
knowing each particular sampling rate
$a_{i,j}$, the enumeration of all the read types is no longer
necessary.\vadjust{\goodbreak} Fortunately, it is possible under
some suitable sampling rate models, including both our uni-\break form model
and insert length model. For example, in
the uniform model, ${\sum_{j=1}^J{a_{i,j}}}=n(l_i-r+1)$,
where $n$ is the total number of mapped
reads, $l_i$ is the length of transcript $i$ and $r$ is the read
length. Similarly,\break in the insert length
model, ${\sum_{j=1}^J{a_{i,j}}}=\break{\sum
_{q(r)>0}nq(r)(l_i-r+1)}$.\vspace*{1pt}

Using this trick, we can take only the read types with at least one
read being mapped and collapse them to
categories $C_1,C_2,\ldots,C_K$. Accordingly, the optimization problem
[equation~(\ref{optimization})] is
reduced to
\begin{eqnarray}\label{optimization.reduce}
&&\mbox{maximize}\quad{n}^T\log({A^T}{\theta})-W^T\theta\nonumber
\\[-8pt]\\[-8pt]
&&\quad\mbox{s.t. } {\theta} \geq0, \nonumber
\end{eqnarray}
where ${n}$ is a $K\times1$ column vector for the collapsed read
counts for categories $C_1,C_2,\ldots,C_K$,
$A$ is a $I\times K$ matrix for the collapsed sampling rates and
${\theta}$ is the isoform abundance vector.
$W$ is a $I\times1$ vector with the $i$th element $W_i$=$\sum
_{j=1}^J{a_{i,j}}$ computed based on the
corresponding sampling rate model.

In a more complex sampling rate model, for example, when $a_{i,j}$
depends on the particular nucleotide sequence of
read $s_j$, the optimization problem [equation~(\ref
{optimization.reduce})] can still be solved. However, all
the read types (including the read types with $n_j=0$) will have to be
enumerated and each sampling rate
$a_{i,j}$ will have to be computed.

%s6 ###
\section{Information Theoretic Analysis}
\label{sec:information}

Many considerations impact the choice of sequencing protocol in an
experimental design. One such choice is
relative cost of sequencing. In this case, the experimentalist may be
interested in choosing the sequencing
protocol (paired end or single end) that provides the best estimate of
isoform abundance at the least relative
cost. This section outlines the statistical argument for why, in
typical situations, paired end sequencing
can produce better estimates of transcript abundance compared to single
end sequencing at a fixed number of
sequenced nucleotides (cost). The theoretical analysis aims to show
that for the same number of sequenced
nucleotides, the Fisher information in the insert length model is more
than double the Fisher information in
the single end read model. Since estimates in RNA-Seq are maximum
likelihood estimators, their variance of
the estimator converges to the reciprocal of the Fisher information.
Thus, larger Fisher information
produces estimators with improved accuracy.

%s6.1 ###
\subsection{Theoretical Analysis}

Consider the following quite
simple example showing the increase in information as the fraction of
reads unique to each isoform grows:

\begin{example}
Continuing Example $1$, suppose\break that isoform 1 and isoform 2 have
Poisson rate
parameters $\theta_1$ and $\theta_2$, respectively, where
$\theta_2=1-\theta_1$ and probability $0<\alpha, \beta< 1$,
respectively, of producing a read unique to the isoform. Let $n_1$ be
the reads unique to $1$,
$n_2$ the reads unique to $2$ and $n_3$ the reads which cannot be
distinguished between the isoforms. Assume there are $n$ total reads
in the sample, and assume there is uniform fragmentation which gives
rise to three
categories:
\begin{eqnarray*}
n_1 &=& \operatorname{Po}(n \alpha\theta_1),\\
n_2 &=& \operatorname{Po}(n \beta\theta_2),\\
n_3 &=& \operatorname{Po}\bigl(n \bigl((1-\alpha)\theta_1+(1-\beta)\theta_2\bigr)\bigr).
\end{eqnarray*}

Fix $\alpha< \beta$ as known and compute the information in this
distribution with $\theta_1$ as the unknown parameter as a function of
$\alpha$ using the definition that the
information is equal to the variance of the derivative of the log
likelihood with respect to $\theta_1$:
\begin{eqnarray*}
I(\theta_1)&=& \operatorname{var}{\biggl(\frac{n_1}{\theta_1}-\frac{n_2}{\theta_2} +
\frac{n_3(\bar{\alpha}-\bar{\beta})}{ \theta_1(\bar{\alpha
}-\bar{\beta})+\bar{\beta}}\biggr)}
\\
&=&n \biggl(\frac{\alpha}{\theta_1}+\frac{\beta}{\theta_2}+\frac{(\bar
{\alpha}-\bar{\beta})}{\theta_1(\bar{\alpha}-\bar{\beta})+\bar
{\beta}} \biggr),
\end{eqnarray*}
where $\bar{x}=1-x$. Thus, $\bar{\alpha}-\bar{\beta}=\beta-\alpha
$, and
$\delta:=\beta-\alpha$ and $\alpha$ are re-parameterizations of
$\beta, \alpha$. The last equation shows that the partial
derivatives of the information with respect to $\alpha$ and with
respect to $\delta$ are positive.
\end{example}

Note that in the example above, no generality is lost by assuming
$\beta>\alpha$ since $\theta_1$ and $\theta_2$ can be interchanged
with no effect on the model.
%%
%Suppose two isoforms for the same gene both have the same probability
%$0<\alpha< 1$ of
%producing a unique read.
%
%Let the Poisson rate parameters for isoform $1$
%be equal to $0<\theta_1<1$ and $\theta_2=1-\theta_1$. Suppose there
%are $n$ total reads in the sample, and let $n_1$ be the reads unique to
%$1$, $n_2$ the reads unique to $2$ and
%$n_3$ the reads which cannot be distinguished between the isoforms. Then
%
%n_1 &=& Po(n \; \alpha\theta_1)\\
%n_2 &=& Po(n \; \alpha\theta_2)\\
%n_3 &=& Po ( n \; ((1-\alpha)\theta_1+(1-\alpha)\theta_2) )\\
%
%Fix $\alpha$ as known and compute the information in this
%distribution with $\theta_1$ as the unknown parameter as a function of
%$\alpha$ using the definition that the
%information is equal to the variance of the derivative of the log
%likelihood with respect to $\theta_1$:
%
%I(\theta_1)&=& var{(\frac{n_1}{\theta_1}-\frac{n_2}{\theta_2})}\\
%&=&n \; \frac{\alpha}{\theta_1}+\frac{\alpha}{\theta_2}\\
%&=&n \; \alpha(\frac{1}{\theta_1}+\frac{1}{\theta_2})
%
%which is increasing in $\alpha$ for any $\theta_1$. Hence, for any
%expression level $\theta_1$, the larger the $\alpha$ or fraction of
%unique reads, the larger the information contained in the density
%parameterized by $\theta_1$.
%
%%
%}

To see that for a fixed cost of sequencing (number of sequenced
nucleotides) the statistical model produced
by paired end sequencing has more information than single end
sequencing, it is necessary to show that the
information obtained by twice as many single end reads in a single end
sequencing experiment is smaller than
that obtained by a paired end sequencing experiment. Such a comparison
necessarily depends on each gene, its
isoforms and their relative abundance. The computation of the Fisher
information for a typical such example
is presented below, and the computation shows that the example easily
generalizes to other configurations of
isoforms.

\begin{example}%[Continuation of Example $\ref{}$]
Continuing the running example, consider reads of length $r=100$ bp and
paired end insert size $x=200$ bp in
the schematic of three exons in Figure \ref{fig:alternative_splicing}, where the length of exons $1$ and $3$
is $500$~bp and exon $2$ is $e=50$~bp (see
Figure~\ref{fig:model_gene}).

%f7 ###
\begin{figure}

\includegraphics{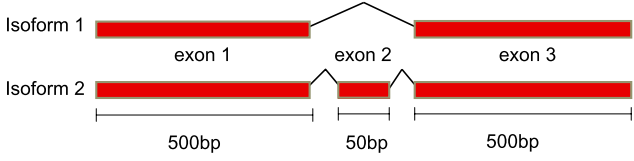}

\caption[A model gene for simulation study]{A model gene for the study of
Fisher information and accuracy of the single end and insert length
models.}\label{fig:model_gene}
\end{figure}

For a single end read experiment, $\alpha_s$ is the probability that a
read includes any part of the included
exon (i.e., uniquely identifies isoform 2), so for the read length of $r$,
\[
\alpha_s=\frac{r-1+e}{1{,}000 + e -r +1}
\]
and $\beta_s$ is the probability that a read includes any part of the
spliced junction (i.e., uniquely identifies isoform~1)~so
\[
\beta_s=\frac{r-1}{1{,}000-r+1}.
\]

For a paired end read experiment, with $x$ the insert length, $\alpha_p$,
the probability that a read uniquely identifies the second isoform, is
\[
\alpha_p= \frac{e+x-1}{1{,}000+e-x+1},
\]
and
$\beta_p$, the probability that a read uniquely identifies the
first isoform, is
\[
\beta_p= \frac{x-1}{1{,}000-x+1}.
\]

For a concrete example, suppose $\theta_1=2\theta_2$. Assume further
that there are twice as many single end
reads (a sample size of $2n$) compared to the $n$ reads in a~paired end run:
\[
I_s:=2n \biggl( \frac{3}{2} {\alpha_s}+3{\beta_s}+\frac{(\bar{\alpha
_s}-\bar{\beta_s})}{(2/3)(\bar{\alpha_s}-\bar{\beta
_s})+\bar{\beta_s}} \biggr),
\]
and the information in a paired end run for a fixed insert size is
\[
I_p:=n\biggl(\frac{3}{2} {\alpha_p}+3{\beta_p}+\frac{(\bar{\alpha
_p}-\bar{\beta_p})}{(2/3)(\bar{\alpha_p}-\bar{\beta
_p})+\bar{\beta_p}} \biggr).
\]

Plugging in numbers $x=200$, $e=50$, and $ r=30$ gives
\[
\frac{I_s}{I_p}= \frac{0.31}{1.12}=0.28.
\]

In other words, in the insert length model, the maximum likelihood
estimator of $\theta_1$ has asymptotic
variance roughly $3$ times larger in the single end read experiment
than in the paired end experiment.

Of course, this ratio will change if the parameters change. For
instance, $I_s/I_p=0.63$ if $x=200$, $e=50$
and $r=70$; $I_s/I_p=0.47$ if $x=200$, $e=100$ and\break $r=50$.
\end{example}

The next section gives simulation results for a related example.

%s6.2 ###
\subsection{Simulation Study}
\label{sim.study}

Simulations were used to study the following questions: (1) the quality
of the
proposed model at estimating isoform-specific gene
expression, especially when the insert length is variable, and (2)
whether abundance estimates from paired end reads are more reliable than
abundance estimates from single end reads.

To address these questions, reads were simulated from a ``hard case''
where a gene has three
exons of lengths 500 bp, 50 bp and 500 bp, respectively (see
Figure~\ref{fig:model_gene}); the middle exon can
be skipped, producing two different isoforms of the gene. Since the
middle exon is short, this case has been shown to be difficult
for isoform-specific gene expression estimation
in~\citet{Jiang2009}.

%f8 ###
\begin{figure*}
\begin{tabular}{@{}cc@{}}

\includegraphics{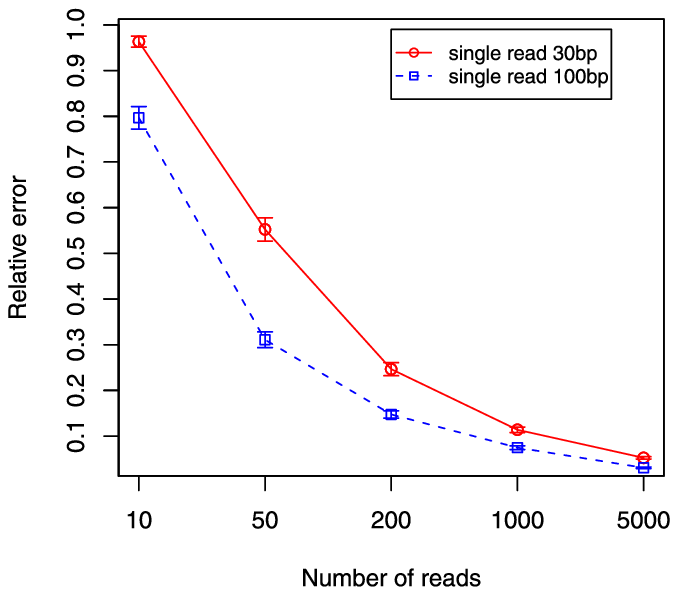}
&\includegraphics{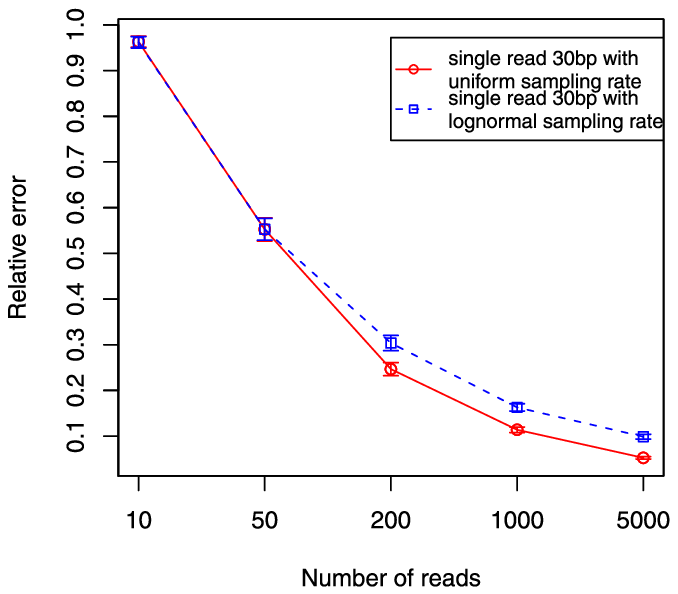}\\
(a)&(b)\\

\includegraphics{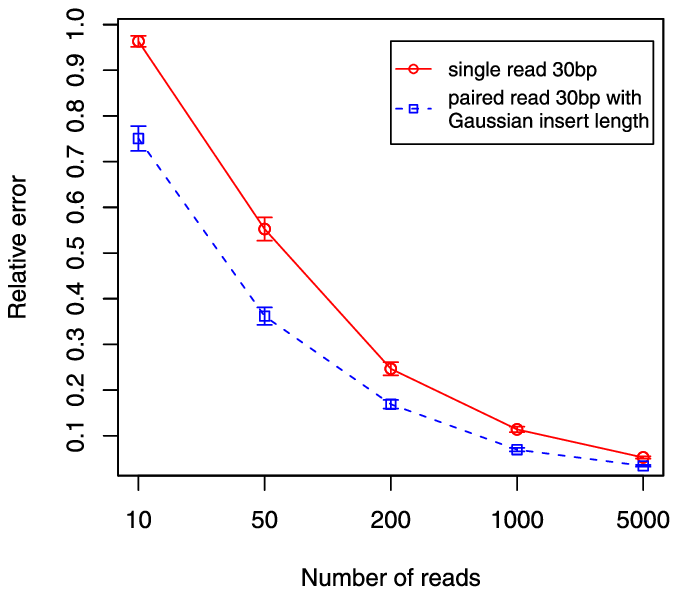}
&\includegraphics{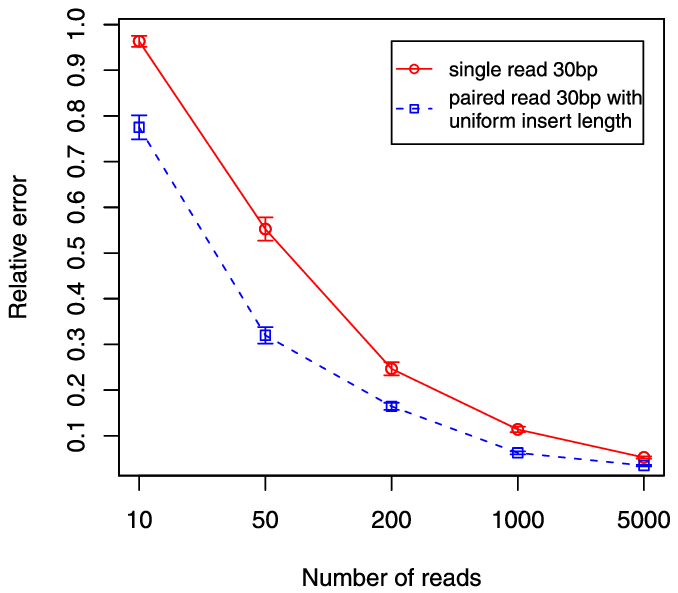}\\
(c)&(d)
\end{tabular}
\caption{Relative error of different read generation models. X axis is
the sample size, that is, the number of
reads that is generated in each simulation experiment. Y axis is the
mean relative error based on $200$
simulation experiments. The error bars give the standard errors of the
sample means. In the figures, single
30 bp reads generated with uniform sampling rate (solid curves) are
compared to (dashed curves) \textup{(a)}~single 100 bp
reads, \textup{(b)} single 30 bp reads generated with lognormal sampling rate,
\textup{(c)} paired end 30 bp reads generated with
Gaussian insert size and \textup{(d)} paired end 30 bp reads generated with
uniform insert size. When compared with $n$
(e.g., $5{,}000$) single end reads, $n/2$ (i.e., $2{,}500$) pairs of paired
end reads were
used.}\label{fig:simulation}
\vspace*{8pt}
\end{figure*}

In the simulation, the two isoforms were assumed to have equal
abundance. Reads were simulated using different models and
parameters described in detail below and estimate isoform
abundances as described in Section~\ref{sec:MLE}. The relative
error of estimation was computed based on the empirical relative
$L^2$ loss:
\begin{eqnarray*}
\frac{\Vert{\theta}-{\hat{\theta}}\Vert_2}{\Vert\theta\Vert_2}&=&\frac{\sqrt
{(\theta_1-\hat{\theta}_1)^2+(\theta_2-\hat{\theta}_2)^2}}{\sqrt
{\theta_1^2+\theta_2^2}}
\\[5pt]
&=&\frac{\sqrt{(1/2-\hat{\theta
}_1)^2+(1/2-\hat{\theta}_2)^2}}{\sqrt{2}/2},
\end{eqnarray*}
where ${\theta}=[\theta_1,\theta_2]=[\frac12,\frac12]$ is the
true isoform abundance vector, and
${\hat{\theta}}=[\hat{\theta}_1,\hat{\theta}_2]$ is the estimated
isoform abundance vector after normalization so that
\mbox{$\hat{\theta}_1+\hat{\theta}_2=1$}. Each simulation experiment
was re-\break peated $200$ times to get the sample mean and standard
error of the relative error.

%s6.2.1 ###
\subsubsection{Simulating single end reads with uniform\break sampling}

To explore the quality of estimation in the uniform sampling approach,
single end reads with length 30 bp
using the uniform sampling model were generated. Five separated
experiments were performed to investigate the
effect of sample size on the estimation procedure using sample sizes of
$10$, $50$, $200$, $1{,}000$ and $5{,}000$,
respectively. The solid curve in Figure~\ref{fig:simulation}(a) gives
the sample mean and standard error of the
relative error. It is clear that relative error decreases as the sample
size increases.

To examine whether longer reads can provide better estimates, all the
simulation experiments were repeated
with read length $100$ bp. Figure~\ref{fig:simulation}(a) shows
the comparison between read lengths of $30$~bp and $100$~bp. As expected,
$100$ bp reads produce smaller error than $30$~bp
reads.

%s6.2.2 ###
\subsubsection{Simulating single end reads with nonuniform sampling}

In real UHTS data, the read distribution is not uniform. To
evaluate how well the RNA-Seq methodology performs in this
regime, simulations were performed where the positions of reads
were sampled from a log-normal distribution. Specifically, up to
a scalar multiple, the true sampling rates $a_{i,j}$ are
independently and identically distributed random variables which
follow\ log-normal distribution with mean $\mu=0$ and standard
deviation $\sigma=1$.

Figure~\ref{fig:simulation}(b) gives the comparison between reads
that were sampled from uniform distribution and reads that were
sampled from log-normal distribution The figure shows that
nonuniform reads produce estimates which appear consistent,
albeit with larger error than with uniform reads.

%s6.2.3 ###
\subsubsection{Simulating paired end reads}
\label{simu_pair}

This section investigates whether, in simulation, paired end reads can
provide more information than single
end reads. When insert lengths do not have a simple distribution,
closed form expressions for the
information are difficult to obtain. Simulation studies are thus
important tools for analyzing such
situations. For this purpose, paired end reads of length $30$ bp with
insert size following a normal
distribution with mean $\mu=200$ bp and standard deviation $\sigma
=20$ bp were generated. For a given insert
size, read pairs were generated using a uniform sampling model.
Figure~\ref{fig:simulation}(c) shows that the
paired end reads produce smaller errors than single end reads with the
same number of sequenced nucleotides:
to make the comparison comparable on the level of total sequenced
bases, $n/2$ pairs of paired end reads were
used when compared with $n$ single end reads.

When the insert size was generated using a uniform distribution, for
example, the effective insert size is uniform
within $200 \pm20$ bp, similar results were produced [see Figure~\ref{fig:simulation}(d)]. Comparing
Figure~\ref{fig:simulation}(d) with Figure~\ref{fig:simulation}(a)
shows that paired end 30 bp reads produce
similarly accurate estimates as 100 bp single end reads, which means
that, on average, paired end reads provide
more information per nucleotide being sequenced.

%s6.2.4 ###
\subsubsection{Simulating with other parameters}

\begin{figure*}
\begin{tabular}{@{}cc@{}}

\includegraphics{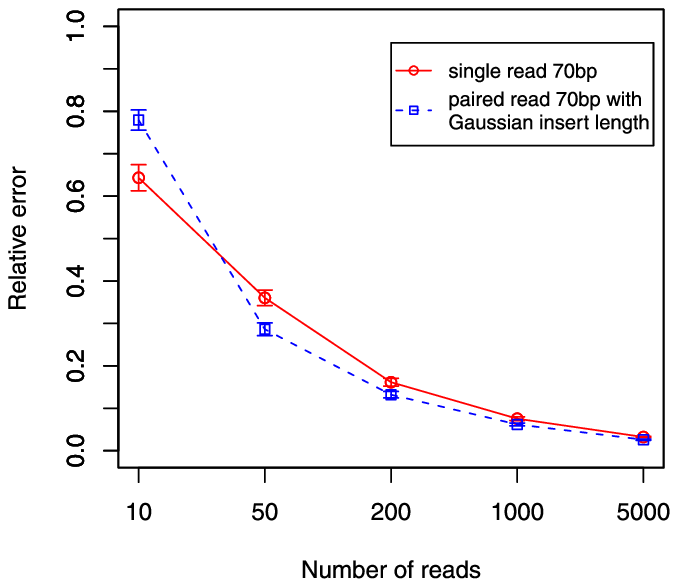}
&\includegraphics{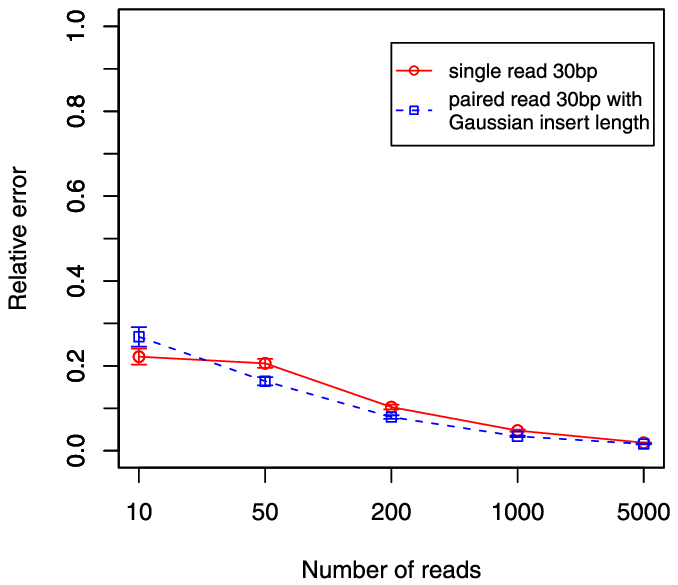}
\\
(a)&(b)\\
\multicolumn{2}{@{}c@{}}{
\includegraphics{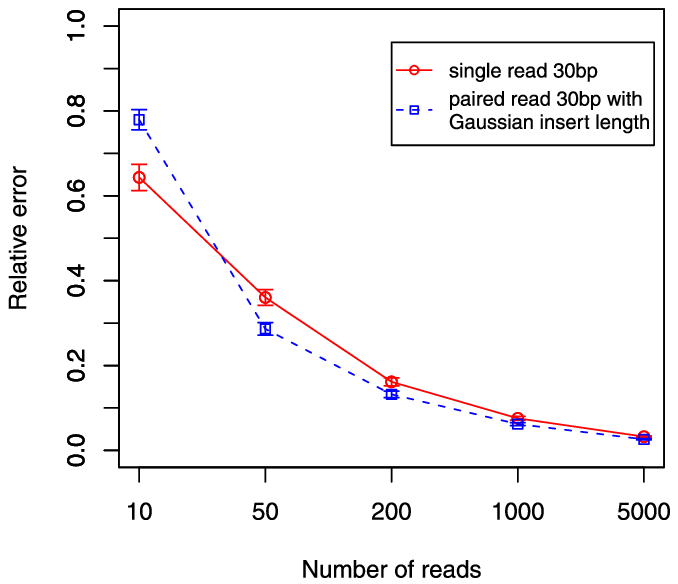}
}\\
\multicolumn{2}{@{}c@{}}{c}
\end{tabular}
\caption{Relative error of single end reads (solid curves) and paired
end reads (dashed curves) with different
settings of parameters: \textup{(a)}~70~bp reads, true isoform expression vector
$(0.5, 0.5)$ \textup{(b)}~30~bp reads, true isoform expression vector $(0.1, 0.9)$
(c) 30 bp reads, true isoform expression vector $(0.5, 0.5)$, exon
lengths (500
bp, 200 bp, 500 bp). When compared with $n$ (e.g., $5{,}000$) single end
reads, $n/2$ (i.e., $2{,}500$) pairs of
paired end reads were used.}\label{fig:others}
\vspace*{6pt}
\end{figure*}

We also performed simulations with other settings of parameters, for
instance, with read length 70 bp, with
true isoform expression vector $(0.1, 0.9)$ or with exon lengths (500
bp, 200 bp, 500 bp). The results are
shown in Figure~\ref{fig:others}. In all these simulations, the
advantage of paired end sequencing over
single end sequencing is obvious for moderate sampling ($50 \leq n \leq
1{,}000$), as in typical cases for sequencing data.

%f9 ###

%s7 ###
\section{Discussion}

The insert length model presented in this paper is a flexible
statistical tool. The model has the capacity
to accommodate oriented reads from Illumina data and to model fragment
specific biases in the probability of
each fragment being sequenced. In~Sec\-tion~\ref{modelassumptions} the
model has been derived when the
experimental step of fragmentation is assumed to be approximated by a
Poisson point process, and a transcript
is assumed to be retained in the sample in proportion to the fraction
of transcripts of its length estimated
after sequencing. These assumptions are at once simplifying and
realistic. As experimental protocols
improve, it is likely they will better model RNA-Seq data.

At the current time, several improvements may be made to the model to
increase its accuracy. First, the
read sampling rate is undoubtedly nonuniform, as it depends on
biochemical properties of the sample and
fragmentation process as experimental studies have highlighted (see
\cite{INGOLIA}; \cite{VEGA}, and
\cite{QUAIL}). This effect becomes more apparent for longer fragments
such as those used in paired end
library preparation. Explicit models for the sampling rates are
difficult to obtain, but doing so is an area
of future research. Recent research (see \cite{Hansen2010}; \cite{Li2010}) has shown that the
nonuniformity can be modeled and estimated quite well from the data. It
may be possible to combine these models
with our approach to improve the estimation performance.

Statistical tests of the reproducibility of the nonuniformity of
reads shows a consistent sequence specific bias across biological
and technical replicates of a~gene. This effect could be due to
bias in RNA fragmentation, bias in other biochemical sample
preparation steps or boundary effects when a gene of fixed length
is fragmented. The last cause of bias can be modeled using Monte
Carlo simulations of a fixed length mRNA sequence subject to a
Poisson fragmentation process and incorporated into the insert length model.

Similarly, the fragmentation and filtering steps ha\-ve not been
explicitly modeled in the insert length model
presented here. Rather, the probability mass function of read lengths,
what is necessary for defining the
model, has been estimated empirically. Improvements to the model could
be made by increasing the precision
of the estimate of the probability mass function of read lengths, for
example, by simulating a fragmentation
and filtering process by Monte Carlo and matching the output of the
simulations to the empirical distribution
function $q(\cdot)$. If such modeling were desired, as described in
Section~\ref{modelassumptions}, the
effects could be easily incorporated into the insert length model. On
the other hand, as experimental
protocols improve, they may reduce this bias and increase the accuracy
of the insert length model as
presented in this paper.

In reality, sequencing mapping is another step that may affect the
analysis. For instance, some reads cannot
be mapped because of sequencing errors and some can be mapped to
multiple places. We have not focused on the
issue of mapping fidelity because we restrict attention to the reads
which did map uniquely. We are also not
taking into account mapping errors which themselves require statistical
modeling. We have chosen not to
model these errors partly because some mapping errors are
platform-dependent (i.e., different sequencing errors
tend to be made by the Illumina vs. other platforms).

In some applications, the parameters of interest to biologists are not
the RPKMs for isoforms 1 and~2, but
rather the relative expression ratio of both isoforms. One way to
estimate the ratio is to reparameterize
the problem with $\theta_1$ as a first parameter and a second
parameter $\mu= \theta_1 /\theta_2$. The
reparameterization will make the model no longer linear in the
parameters, therefore harder to solve. Also,
the choice of $\mu$ is not straightforward when there are 3 or more
isoforms. An easier way is to estimate
$\mu$ indirectly after estimating $\theta_1$ and~$\theta_2$.

We believe that technological improvements that produce longer read
lengths will not diminish the relevance
of the insert length model. Paired end models will be relevant at least
until read lengths are comparable to
the length of each transcript, and perhaps longer for reasons of cost.
Since many transcripts are larger than
$10^{4}$ nucleotides, and longer in some important cases, such a time
is unlikely to occur in the next few
years. Further, longer insert lengths and reads combined with the
insert length model in this paper will aid
in discrimination of complex isoforms and estimation of
isoform-specific poly-A tail lengths. Thus, we do not
foresee any imminent obsolescence of this model.

%variety of lengths. Longer read length or longer
%insert sizes may introduce biases in quantifying short transcripts.
%Therefore, several experiments with
%varying insert lengths or read lengths could be performed to allow for
%different sets of transcripts to be
%targeted in each experiment. Despite this, longer insert lengths do
%help in discovery of novel transcripts
%and determining the correlation between distant splicing events.
%}
While the model developed in this paper has the potential for great use
and extends current methodology for isoform-specific estimation,
the model assumes that the complete set of isoforms of a gene have been
annotated. De novo discovery of isoforms from a sample is an important
and difficult statistical problem that we have not addressed in this
paper.
Another shortcoming of the model is that in order for statistical
inference to be accurate, with the current
short read technology, the number of
isoforms should be relatively small (e.g., 2--5). We expect these
challenges to motivate methodological development in the field of
RNA-Seq in
the coming years.

% a genome-wide test of read
%uniformity using a modified Kolmogorov-Smirnov test was performed
%on a per gene basis. This test is as follows: for each of two
%biological replicates, all reads are aligned to a single sequence
%which consists of concatenation of all exons from a gene. The
%reads are then mapped and their positions are convolved with
%independent uniform $(0,1)$ random variables. This produces two
%empirical distributions $\hat{F}_i$ for $=1,2$ according to
%whether the reads are derived from the first or second biological
%replicate. The exact null distribution for the KS test is
%computed for genes with few numbers of reads. The set of
%$p$-values resulting from this test are plotted in
%Figure~\ref{fig:nonuniformity-reproducible}. Despite a small peak
%near $0$, the uniformity of the p-values suggests that sampling
%rates of transcripts in these experiments are consistent with
%each other; this includes rates of at which PCR produces a clonal
%population of identical molecules.

%%
%
%modified K-S tests of read positions in each gene between two
%biological replicates of the first
%sample.\label{fig:nonuniformity-reproducible}}
%
%
%}

In conclusion, this paper has presented a statistical model for RNA-Seq
experiments which provides estimates
for isoform specific expression. Finding such estimates is difficult
using microarray technology, focusing
interest in UHTS to address this question. In addition to modeling, the
paper has presented an in-depth
statistical analysis. By using the classical statistical concept of
minimal sufficiency, a computationally
feasible solution to isoform estimation in RNA-Seq is provided.
Further, statistical analysis quantifies the
perceived gain in experimental efficiency from using paired end rather
than single end read data to provide
reliable isoform specific gene expression estimates. To the best of our
knowledge, this is the first
statistical model for answering this question.

%say a forthcoming paper introduces other tests?

\section*{Acknowledgments}
The authors would like to acknowledge Jamie Geier Bates for providing
the data used in Section
\ref{applications}, and Patrick O. Brown for useful discussions, as well
as several anonymous referees whose comments impro\-ved the clarity of the
manuscript. We thank Michael
Saunders for his help in interfacing the PDCO package. Salzman's
research was supported in part by NSF Grant DMS-08-05157. Jiang's
research was supported in part by NIH Grant 2P01-HG000205. Wong is funded
by a NIH Grant
R01-HG004634. The computation in
this project was performed on a system supported by NSF computing
infrastructure Grant DMS-08-21823.

% imsref loaded by arune.pranskunaite, 2011-01-31 13:17:57
%

\end{document}